\documentclass[preprint,sort&compress,12pt]{elsarticle}

\usepackage[top=1in, bottom=1in, left=1in, right=1in]{geometry}

\usepackage{graphicx}
\usepackage{amsmath}

\usepackage{natbib}
\usepackage{amssymb}

\usepackage{amsthm}

\usepackage{lineno}
\usepackage{subfig}
\usepackage{enumerate}
\usepackage{fullpage}
\usepackage{algorithm}
\usepackage{algpseudocode}
\usepackage[colorinlistoftodos]{todonotes}

\usepackage{hyperref}
\biboptions{sort,compress} 

\usepackage{xcolor}

\newcommand*\patchAmsMathEnvironmentForLineno[1]{%
  \expandafter\let\csname old#1\expandafter\endcsname\csname #1\endcsname
  \expandafter\let\csname oldend#1\expandafter\endcsname\csname end#1\endcsname
  \renewenvironment{#1}%
     {\linenomath\csname old#1\endcsname}%
     {\csname oldend#1\endcsname\endlinenomath}}% 
\newcommand*\patchBothAmsMathEnvironmentsForLineno[1]{%
  \patchAmsMathEnvironmentForLineno{#1}%
  \patchAmsMathEnvironmentForLineno{#1*}}%
\AtBeginDocument{%
\patchBothAmsMathEnvironmentsForLineno{equation}%
\patchBothAmsMathEnvironmentsForLineno{align}%
\patchBothAmsMathEnvironmentsForLineno{flalign}%
\patchBothAmsMathEnvironmentsForLineno{alignat}%
\patchBothAmsMathEnvironmentsForLineno{gather}%
\patchBothAmsMathEnvironmentsForLineno{multline}%
}
\usepackage{color,soul}
\usepackage{enumitem}
\usepackage{mathtools}
\usepackage{booktabs}

\definecolor{lightblue}{rgb}{.90,.95,1}
\definecolor{darkgreen}{rgb}{0,.5,0.5}

\definecolor{lightgreen}{rgb}{.90,1,0.90}

\newcommand{\bstau}{\boldsymbol{\tau}}

\newcommand{\bstaurans}{\tilde{\boldsymbol{\tau}}^{rans}}

\newcommand{\bs}[1]{\boldsymbol{#1}}

\usepackage{longtable}
\usepackage{tabularx,ragged2e,booktabs,caption}
\newcommand{\RN}[1]{%
  \textup{\uppercase\expandafter{\romannumeral#1}}%
}

\usepackage{bibentry}
\newcommand{\ignore}[1]{}
\newcommand{\nobibentry}[1]{{\let\nocite\ignore\bibentry{#1}}}

\journal{XXX}

\begin{document}

\begin{frontmatter}

%\tableofcontents
% \listoftodos


\title{Incorporating Prior Knowledge for Quantifying and Reducing Model-Form Uncertainty in RANS Simulations}



\author{Jian-Xun Wang\corref{corjxw}}
\author{Jin-Long Wu\corref{corjl}}
\author{Heng Xiao\corref{corxh}}
\cortext[corxh]{Corresponding author. Tel: +1 540 231 0926}
\ead{hengxiao@vt.edu}

\address{Department of Aerospace and Ocean Engineering, Virginia Tech, Blacksburg, VA 24060, United States}

\begin{abstract}
Simulations based on Reynolds-Averaged Navier--Stokes (RANS) models have been used to support 
high-consequence decisions related to turbulent flows. Apart from the deterministic model predictions, 
the decision makers are often equally concerned about the predictions confidence. Among the uncertainties 
in RANS simulations, the model-form uncertainty is an important or even a dominant source. Therefore, 
quantifying and reducing the model-form uncertainties in RANS simulations are of critical importance to 
make risk-informed decisions. Researchers in statistics communities have made efforts on this issue by 
considering numerical models as black boxes. However, this physics-neutral approach is not a most 
efficient use of data, and is not practical for most engineering problems. Recently, we proposed an 
open-box, Bayesian framework for quantifying and reducing model-form uncertainties in RANS simulations 
based on observation data and physics-prior knowledge. It can incorporate the information from the 
vast body of existing empirical knowledge with mathematical rigor, which enables a more efficient usage of 
data. In this work, we examine the merits of incorporating various types of prior knowledge in the 
uncertainties quantification and reduction in RANS simulations. The result demonstrates that informative 
physics-based prior plays an important role in improving the quantification of model-form uncertainties, 
particularly when the observation data are limited. Moreover, it suggests that the proposed Bayesian 
framework is an effective way to incorporate empirical knowledge from various sources of turbulence modeling. 
  
\end{abstract}

\end{frontmatter}

\section{Introduction}
\label{sec:intro}

In recent years, Computational Fluid Dynamics (CFD) models based on Reynolds-Averaged
Navier--Stokes (RANS) equations have been used to support high-consequence decisions related to
turbulent flows. For example, RANS simulations have been used to support the development of the emergency
evacuation plans in scenarios of pollutant releases in cities~\cite{nasstrom2005}. In the nuclear energy
industry, government regulators and plant operators have explored the feasibility of using RANS
models to model the thermo-hydraulic systems in nuclear power plants to support safety
assessments and licensing~\cite{bieder03,Scheuerer:2005cu}. In these contexts, quantifying and
reducing uncertainties in the RANS simulations is of critical importance for the stake-holders to
make risk-informed decisions.

Although the uncertainties in RANS simulations may also occur due to uncertain inputs and parameters, meshes,
and numerical methods, the uncertainties that are attributed to the inadequacy and
intrinsic assumptions in the physical model of turbulence, referred to as model-form uncertainties,
are most challenging to quantify in RANS simulations. Oliver et
al.~\cite{oliver2009uncertainty} was the first to propose a ``composite model theory'' in the
context of RANS modeling, where the model-form uncertainties are localized to the Reynolds stress
tensors. Dow and
Wang~\cite{dow11quanti} introduced uncertainties in the turbulent eddy viscosity to estimate the
structural uncertainties in the $k$--$\omega$ model~\cite{wilcox1998turbulence}. On the other hand,
Iaccarino and co-workers~\cite{emory2011modeling,gorle2013framework,gorle2014deviation,
emory2013modeling,emory14estimate} introduced uncertainties
directly to the Reynolds stress, and estimated the RANS modeling uncertainties by perturbing the predicted
Reynolds stress in a physically realizable range.

Recently, Xiao et al.~\cite{xiao-mfu} proposed a Bayesian framework for quantifying and reducing
uncertainties in RANS simulations by incorporating velocity observation data and physics-based prior
knowledge. A key element in this framework is the inference of
Reynolds stress discrepancy field from sparse velocity observations. Therefore, it can be considered a
Bayesian calibration framework. This is in contrast to Oliver et al.~\cite{oliver2009uncertainty}
and Iaccarino and co-workers~\cite{emory2011modeling}, who focused solely on the forward propagation
of uncertainties in Reynolds stresses to velocity and other Quantities of Interests (QoIs).
On the other hand, the physics-neutral Bayesian framework of Kennedy and O'Hagan~\cite{kennedy2001bayesian} 
treats numerical models as black boxes and does not allow straightforward representation of the prior 
knowledge~\cite{brynjarsdottir2014learning}. In addition, the available data are usually too sparse in engineering 
application to drive such physics-neutral approach, which poses a practical hurdle for
its acceptance in engineering communities.  Compared to the Bayesian model calibration framework of Kennedy and
O'Hagan, a notable feature of the method of Xiao et al.~\cite{xiao-mfu}
is that it incorporates physics-based prior knowledge, which can effectively augment the information provided by 
the limited amount of data. Sources of prior knowledge range from constraints that can be
expressed with mathematical rigor (e.g., physical realizability of Reynolds stress tensor) to
imprecise, subjective beliefs that are not amenable to mathematical representation (e.g., empirical
knowledge accumulated in the turbulence modeling from decades of applications of RANS simulations). 

The target application scenario of the framework proposed in~\cite{xiao-mfu} is the prediction of
complex turbulent flows with a limited amount of observation data. The framework enables predictions
with quantified uncertainties by combining numerical model simulations and observation data (e.g.,
real-time measurements from sensors in a nuclear power plant or data from air quality monitoring
stations in a city). As in other Bayesian inference frameworks, the posterior distributions of the
predicted QoIs depend on both prior information and data. Since the amount of data is limited in the application
scenario, the effects of prior knowledge are expected to be particularly important. 
The objective of this work is to examine the role of incorporating prior knowledge in
quantifying and reducing uncertainties in RANS simulations. Meanwhile, we will demonstrate that 
physics-based prior knowledge can be expressed and incorporated properly within the proposed 
framework~\cite{xiao-mfu}. 

In the context of RANS modeling, following types of prior knowledge are considered:
\begin{enumerate}
\item the assumption that Reynolds stress is the dominant source of uncertainty in RANS equations,
\item physical realizability constraints on Reynolds stress tensors,
\item the symmetry (if any) of the flow of concern,
\item  smooth spatial distribution of the Reynolds stresses,
\item  overall understanding on the coherent structures of the flow,
\item subjective belief on the discrepancies of predicted Reynolds stress tensors (or more precisely
  their projections, including the magnitude, shape, and orientation) in different flow regions.
\end{enumerate}
Items 1--2 are strict constraints resulting from the assumption of the modeling and the physical realizability,
which can be guaranteed straightforwardly in the uncertainty
quantification framework~\cite{emory2011modeling}. In contrast, items 3--6 involve analysts' physical
understanding and subjective judgment of the flow. The strict constraints are built
into the framework of Xiao et al.\cite{xiao-mfu}, while proper means are provided for the analyst
to specify subjective priors of various precisions. In this study, we show that 
the inference uncertainties (in Reynolds stresses and other QoIs) are significantly
reduced by using realistic, informative priors compared to using non-informative priors.  We also show that
using more informative priors can lead to similar inference uncertainty reduction compared to that
obtained by incorporating additional observation data. Both comparisons demonstrate that the
empirical knowledge accumulated in the turbulence modeling community can be effectively used for quantifying and
reducing RANS model uncertainties. Although the importance of prior knowledge is demonstrated
specifically in the context of RANS modeling of turbulent flows, it is expected that the conclusions
can be extended to other complex physical systems as well.

The remaining of the paper is organized as follows. In Section~2 the framework introduced in
\cite{xiao-mfu} is summarized with emphasis on the utilization and representation of prior
knowledge. Numerical simulations based on the flow in a channel with periodic constrictions are
presented in Section~3 to highlight the importance of informative prior knowledge. In Section~4 we
further discuss the mapping of Reynolds stress and mean velocities to show the success and
limitation of the framework. Finally, Section 5 concludes the paper.

\section{Uncertainty Quantification Framework and Representation of Prior Knowledge}
\label{sec:method}

\begin{table}[htbp]
  \centering
  \begin{tabular}[c]{|m{0.02\textwidth}|m{0.4\textwidth}|m{0.35\textwidth}|m{0.15\textwidth}|}
    \hline
    & Prior knowledge & Representation in the framework~\cite{xiao-mfu}
    & Demonstrated in cases
    \\
    \hline \hline
    1 &    Reynolds stress is a dominant uncertainty source in RANS equations & modeling of Reynolds stress as
    a random field (built into framework) &   --
    \\     
    \hline
    2  &     physical realizability of Reynolds stresses & parameterization of physical variables, e.g., magnitude and shape of Reynolds stresses  (built into framework) & --
    \\
    \hline
    3 &   symmetry of the flow  & use  of symmetric  basis functions in parameterization  & --
    \\
    \hline
    4 & Reynolds stresses have smooth spatial distribution & use of smooth basis functions in parameterization  &  S1/S2 \\
    \hline
    5 & relative discrepancies of Reynolds stresses in different regions & proper choice of spatially varying variance
     fields $\sigma(x)$ and length scale field $l(x)$ in constructing of a non-stationary Guassian kernel & D1
    \\
    \hline
    6 &  overall knowledge of the flow & proper design of observation locations (experimental design) &  D2
    \\
    \hline
  \end{tabular}
  \caption{Representation of physics-based prior knowledge in the open-box uncertainty
    quantification and reduction framework for RANS modeling proposed in~\cite{xiao-mfu}}
  \label{tab:prior}
\end{table}

The framework for quantifying and reducing model-form uncertainties in RANS simulations as proposed
by Xiao et al.~\cite{xiao-mfu} is summarized below, emphasizing the
 prior knowledge in RANS simulations and their representation in the
framework. The prior knowledge that is incorporated in the framework is presented
in Table~\ref{tab:prior}.

\begin{figure}[!htbp]
  \centering
   \subfloat[Barycentric coordinate]
   {\includegraphics[width=0.5\textwidth]{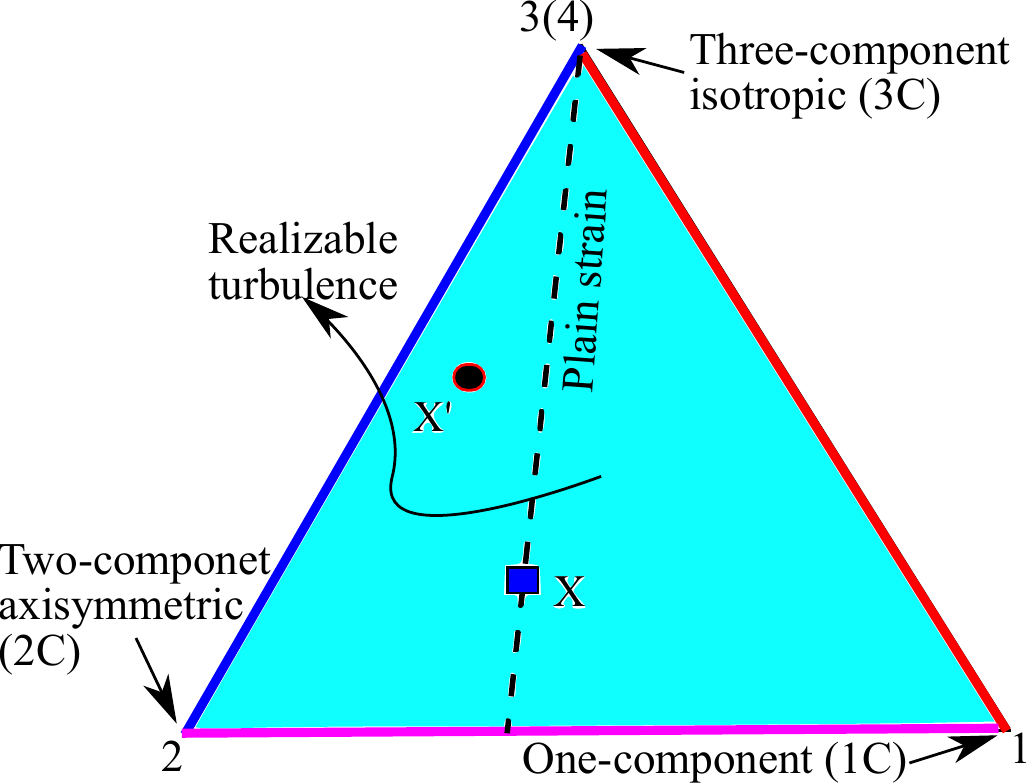}} \hspace{2em}
   \subfloat[natural coordinate]
   {\includegraphics[width=0.4\textwidth]{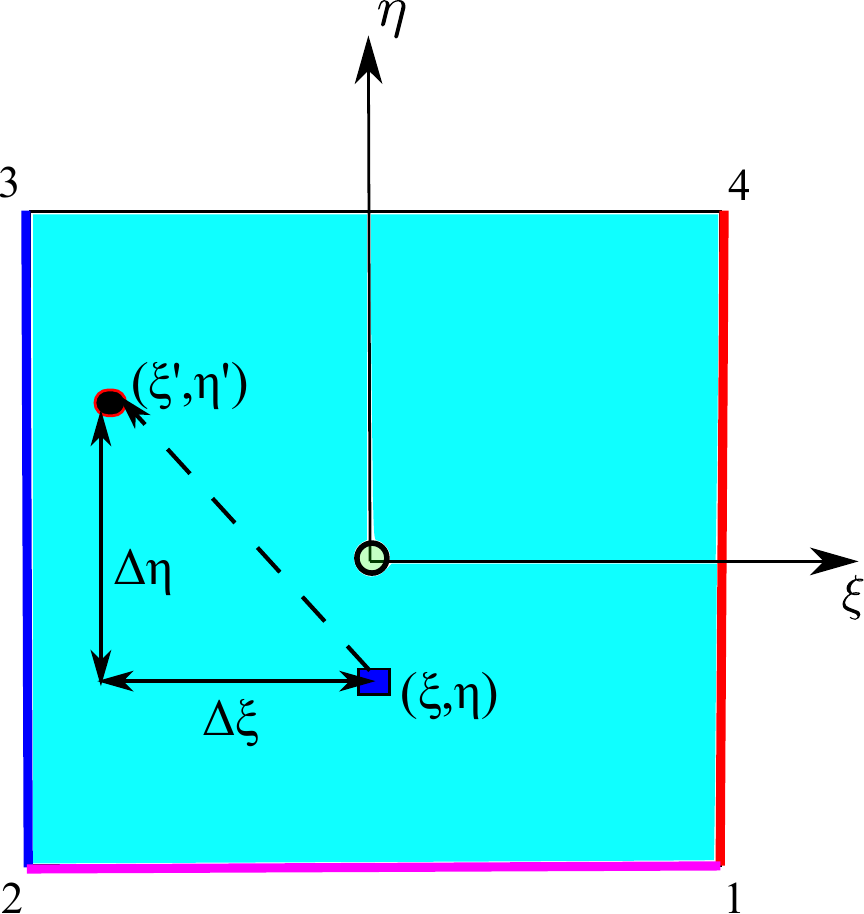}}\\
   \caption{Mapping between the Barycentric coordinate to the natural coordinate, transforming the
     Barycentric triangle enclosing all physically realizable
     states~\cite{banerjee2007presentation,emory2013modeling} to a square via standard finite
     element shape functions. Corresponding edges in the two
     coordinates are indicated with matching colors.}
  \label{fig:bary}
\end{figure}

It is a consensus in the turbulence modeling community that discrepancy in the modeled Reynolds stress
field is the main source of model-form uncertainties in the RANS equations~\cite{oliver2011bayesian}. 
Consequently, uncertainties are injected to the Reynolds stress field by perturbing the
RANS-modeled Reynolds stress (see Prior 1 in Table~\ref{tab:prior}). More precisely, the true
Reynolds stress $\bstau(x)$ is modeled as a random field with the spatial coordinate $x$ as index
and the RANS-predicted Reynolds stress $\bstaurans(x)$ as prior mean.  The Reynolds stress at any
location is a symmetric tensor. Note that an arbitrary perturbation can lead to a tensor that does
not correspond to any physically possible states. Therefore, perturbations are introduced 
within the realizable states of Reynolds stress based on the
physical meaningful projections of the Reynolds stress tensor and not on its individual
components. This is to ensure that all realizations of the random field $\bstau(x)$ are physically
possible (see Prior 2 in Table~\ref{tab:prior}). Specifically, the
Reynolds stress tensor is transformed to physically interpretable variables as follows:

\begin{equation}
  \label{eq:tau-decomp}
  \boldsymbol{\tau} = 2 k \left( \frac{1}{3} \mathbf{I} +  \mathbf{a} \right)
  = 2 k \left( \frac{1}{3} \mathbf{I} + \mathbf{V} \Lambda \mathbf{V}^T \right)
\end{equation}
where $k$ is the turbulent kinetic energy, which indicates the magnitude of $\bstau$; $\mathbf{I}$
is the second order identity tensor; $\mathbf{a}$ is the anisotropy tensor; 
$\mathbf{V} = [\mathbf{v}_1, \mathbf{v}_2, \mathbf{v}_3]$ and 
$\Lambda = \textrm{diag}[\lambda_1, \lambda_2, \lambda_3]$ with
$\lambda_1+\lambda_2+\lambda_3=0$ are the orthonormal eigenvectors and eigenvalues of $\mathbf{a}$,
respectively, indicating the shape and orientation of $\bstau$. The eigenvalues $\lambda_1$,
$\lambda_2$, and $\lambda_3$ are mapped to a Barycentric coordinate $(C_1, C_2, C_3)$ with $C_1 +
C_2 + C_3 = 1$. Consequently, all physically realizable states are enclosed in the Barycentric
triangle shown in Fig.~\ref{fig:bary}a. To facilitate the parameterization, the Barycentric
coordinate is further transformed to the natural coordinate $(\xi, \eta)$, with the triangle mapped
to a square as shown in Fig.~\ref{fig:bary}b.  Finally, uncertainties are introduced to the mapped
quantities $k$, $\xi$, and $\eta$ by adding discrepancy terms to the corresponding RANS predictions,
i.e.,
\begin{subequations}
    \label{eq:delta-def}
  \begin{alignat}{2}
    \log k(x) & = &\ \log \tilde{k}^{rans}(x)  & + \delta^k(x)  \label{eq:kdelta} \\
    \xi (x) & = &\ \tilde{\xi}^{rans}(x) & + \delta^\xi(x)  \\
    \eta(x) & = &\ \tilde{\eta}^{rans}(x) & + \delta^\eta(x)
  \end{alignat}
\end{subequations}
Uncertainties are not introduced to the orientation ($\mathbf{v}_1, \mathbf{v}_2, \mathbf{v}_3$) of
the Reynolds stress. This is to avoid instability caused by possible reverse diffusions.

The smooth spatial distribution of the Reynolds stress is another constraint. The smoothness is
guaranteed by constructing the Reynolds stress discrepancy fields $\delta^k$, $\delta^\xi$, and
$\delta^\eta$ (denoted as $\delta$ generically below) with smooth basis functions (see Priors 3 and 4 in
Table~\ref{tab:prior}).  Specifically, the prior distributions of the discrepancies are chosen as
non-stationary zero-mean Gaussian random fields $\mathcal{GP}(0, K)$ (also know as Gaussian
processes), where
  \begin{equation}
  \label{eq:gp-kernel}
  K(x, x') = \sigma(x)  \sigma(x') 
  \exp \left( - \frac{|x - x'|^2}{l^2}  \right)
\end{equation}
is the kernel indicating the covariance at two locations $x$ and $x'$. The variance $\sigma(x)$ can be 
specified as a spatially varying field (see the contour in Fig.~\ref{fig:domain_pehill} for example) to
reflect the prior knowledge on relative discrepancies of modeled Reynolds stress in different flow regions 
(see Prior 5 in Table~\ref{tab:prior}).  The correlation length scale $l$ can be specified based on 
the local turbulence length scale, but is taken as a constant in this work for simplicity.

The basis set is chosen as the eigenfunctions of the kernel $K$ computed from the
Fredholm integral~\cite{le2010spectral}. This choice of basis function leads to the Karhunen--Loeve
(KL) expansion of the random field. That is, the discrepancy fields $\delta$ can be represented as follows~\cite{xiao-mfu}:

\begin{equation}
  \label{eq:delta-proj}
  \delta(x, \theta) = \sum_{i=1}^\infty \omega_{i} |_{\theta} \; \phi_i (x) , 
\end{equation}
where the coefficients $\omega_{i}$ (denoting $\omega_{k, i}$, $\omega_{\xi, i}$, $\omega_{\eta, i}$)
are independent standard Gaussian random variables depending on the realization of $\theta$. 
In practice the infinite series is truncated to $m$ terms, with $m$
depending on the smoothness of the kernel~$K$.

With the decomposition above, the true Reynolds stress field is modeled as the random
field with RANS modeled Reynolds stress as prior mean and the discrepancies parameterized by the
coefficients $\omega_{k, i}, \, \omega_{\xi, i}, \, \omega_{\eta, i}$ with $i = 1, 2, \cdots, m$.  The
uncertainty distributions of the coefficients are then inferred by using an iterative ensemble
Kalman method, which is an approximate Bayesian inference method widely used for data assimilation
in geoscience communities~\cite{evensen2009data}. In this method the prior distribution of Reynolds
stresses (as parameterized by the coefficients) is first represented by samples drawn from the
prior distribution.  The collection of samples, referred to as prior ensemble, is propagated to velocities
by using a forward RANS solver \texttt{tauFoam}, which computes velocities from a given
Reynolds stress field. This forward RANS solver is developed based
on a conventional steady-state RANS solver in OpenFOAM~\cite{openfoam} by replacing the turbulence
modeling component (i.e., solution of the transport equations for turbulence quantities) with a
supplied Reynolds stress field. With the predicted ensemble of the forward RANS solver, 
the Kalman filtering procedure is used to incorporate velocity observation
data to the prediction, yielding a corrected ensemble.  The procedure is repeated until statistical
convergence is achieved. The converged posterior ensemble is a sample-based representation of the
uncertainty distribution of the Reynolds stresses and other quantities of interests given the
observation data.  For the scenarios as mentioned above, the amount of data used in the inference procedure is limited.
Therefore, proper arrangement of observation locations (i.e., experimental design) is important.
The prior knowledge on the overall feature of the flow can be incorporated as a guidance to conduct the experimental
design, which leads to a more effective use of data. (see Prior 6 in Table~\ref{tab:prior}).
 
\section{Numerical Simulations}
\label{sec:num}
\subsection{Problem Setup}
\label{sec:setup}
\begin{figure}[htbp]
  \centering
  \includegraphics[width=0.75\textwidth]{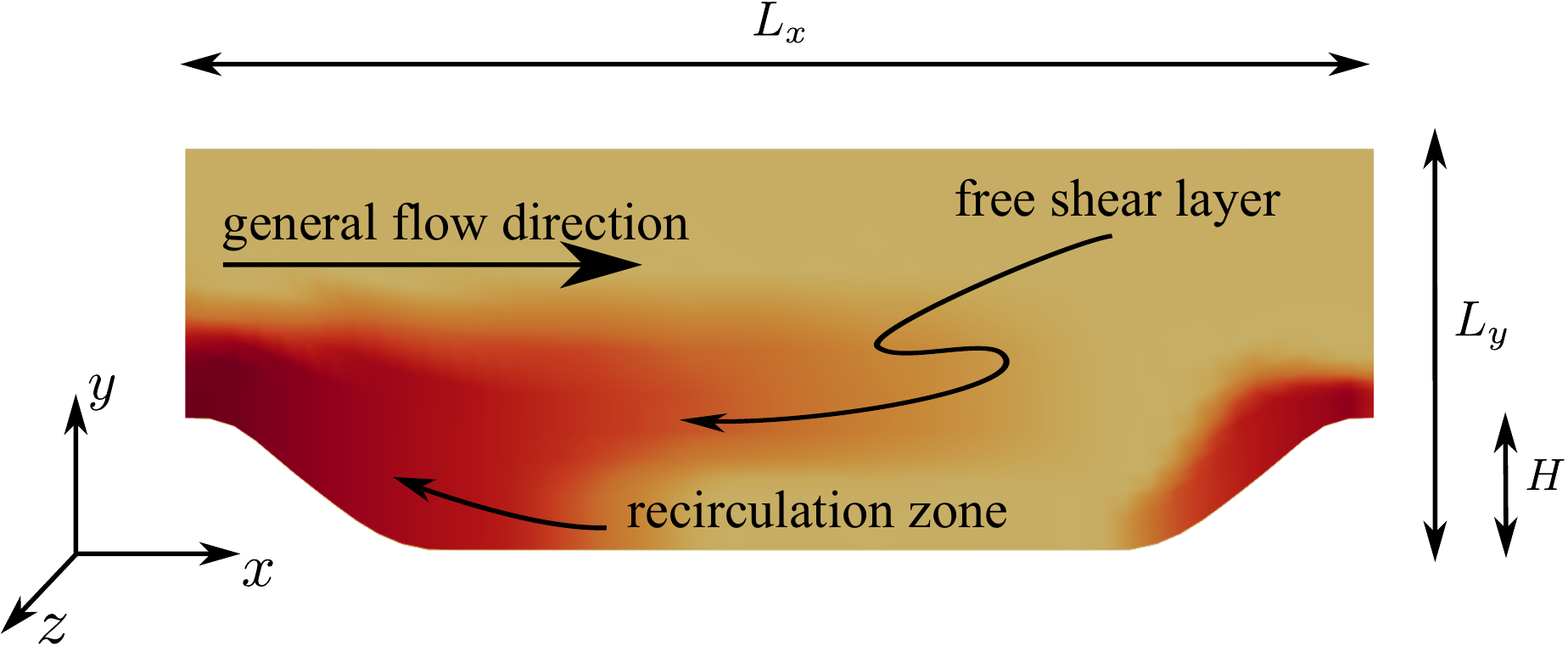}
  \caption{Domain shape for the flow in the channel with periodic hills. 
  The $x$-, $y$- and $z$-coordinates are
  aligned with the streamwise, wall-normal, and spanwise directions, respectively. All dimensions are
  normalized by $H$ with $L_x/H=9$ and $L_y/H=3.036$. The contour shows the variance field
  $\sigma(x)$, where darker color represents larger variance.}
  \label{fig:domain_pehill}
\end{figure}

The flow over periodic hills at Reynolds number $Re_b = 2800$ is 
studied to demonstrate the merits of incorporating physics-based prior knowledge in quantifying
and reducing the model-form uncertainties in RANS simulations. 
Direct numerical simulation (DNS) data~\cite{breuer2009flow} are adopted as the 
benchmark for comparison. The computational domain is shown in Fig.~\ref{fig:domain_pehill}, where all 
dimensions are normalized with the crest height $H$. The Reynolds number $Re_b$ is based on the
height $H$ and bulk velocity $U_b$ at the crest. 

In addition to the DNS benchmark data, synthetic benchmark data are also utilized in this study, 
which is created from the RANS 
model with a specified Reynolds stresses perturbation. The reason of using synthetic truth is 
that we can control the dimensions of the uncertainty space and thus
are able to explore its effects upon the result. Moreover, 
the coefficients vector $\bs{\omega}$ is known from the synthetic truth, which can be used to verify 
the proposed inversion scheme.
To generate the synthetic truth, we use the standard RANS simulation outputs as baseline
and perturb the Reynolds stress field with specified $\delta^{\xi}$, $\delta^{\eta}$, and $\delta^k$. 
For simplicity, we only perturb the $\eta$ with $\delta^{\eta}$ constructed from two KL modes with
coefficients $\hat{\bs{\omega}}_{\eta} = [\hat{\omega}_{\eta, 1}, \hat{\omega}_{\eta, 2}]$, where
$\hat{\omega}_{\eta, 1} = 1.5$, $\hat{\omega}_{\eta, 2} = 1.0$, and $\hat{\cdot}$ denotes  synthetic truth.
Notice that the fundamental difference between the synthetic truth and DNS benchmark is whether the uncertainty
space spanned by the prior ensemble covers the true Reynolds stresses. For synthetic truth, the prior can
cover the truth since we know where it resides. However, the uncertainty space does
not necessarily cover the true Reynolds stresses that generate DNS velocity benchmark. 
This is because the orientations of the Reynolds stresses are not perturbed and that the KL modes
are truncated.   

The observation data are obtained by sparsely observing the velocity fields of the 
truth. Independent and identically distributed Gaussian random noises with 
standard deviation $\sigma_{obs}$ are added to represent the observation
errors. The observations generated from synthetic truth are referred to as ``synthetic observation'', while
those observed from DNS data are denoted as ``DNS observation''. In the following, four
cases are studied to investigate the effects of incorporating the prior knowledge into the 
framework. In Sec.~\ref{sec:complex}, two cases S1 and S2 are studied, where 
synthetic observation data are adopted
to explore the effects of prior knowledge on the dimension of uncertainty space. 
We span a less informative searching space (i.e., prior uncertainty space with a high dimension) in case S1, 
while we span a more informative searching space (i.e., prior uncertainty space with a low dimension) in 
case S2. Note that the prior uncertainty spaces cover the synthetic truth in both cases.
In Sec.~\ref{sec:variance}, two cases D1 and D2 with DNS observations are studied to investigate the
merits of physics-based prior of the variance field $\sigma(x)$ (see Prior 5 in Table~\ref{tab:prior}).
The cases D1 and D2 are performed with the same parameters and setup except for different variance fields. 
In case D2, the empirical knowledge on the $\sigma(x)$ fields, shown
in Fig.~\ref{fig:domain_pehill}, are incorporated into the framework, 
which is in contrast to a uniform $\sigma(x)$ field in the case D1.  
In addition, two scenarios of case D2 with different
arrangements of the observation locations are investigated to demonstrate the merits of physics-based
knowledge on experimental design in Sec.~\ref{sec:experi} (also see Prior 6 in Table~\ref{tab:prior}).  
The mesh and computational parameters used in the uncertainty 
quantification and reduction procedure are summarized in Table~\ref{tab:paraDA}. 
The uncertainties in Reynolds stress anisotropy ($\xi$ and $\eta$) and amplitude ($k$) are
all considered in cases D1 and D2.
The correlation length scale $l$ is chosen based on the length scale of the flow, 
which can be determined from the physical understanding of the flow. In synthetic cases, 
the length scale $l$ is larger to achieve a lower dimensional uncertainty searching space.
For each case, a prior ensemble of 60 samples is employed. 

\begin{table}[htbp]
  \centering
  \caption{Mesh and computational parameters used in the flow over periodic hills.
    \label{tab:paraDA}
  }
    \begin{tabular}[b]{c|c |c |c |c}
      \hline
      cases & S1 & S2 &  D1 & D2 \\
      \hline
      fields with uncertainty  & $\xi$, $\eta$, $k$ & $\eta$ &\multicolumn{2}{c}{$\xi$, $\eta$, $k$}\\
      number of modes  & 18 & 2 &\multicolumn{2}{c}{48}\\
      \hline
      correlation length scale $l/H$$^{\mathrm{(a)}}$  & \multicolumn{2}{c|}{5} & \multicolumn{2}{c}{1}\\
      number of observations  & \multicolumn{2}{c|}{16} & \multicolumn{2}{c}{18}\\
      \hline
      variance field $\sigma(x)$ & \multicolumn{3}{c|}{Non-informative ($\sigma(x)$ = constant)} & Informative$^{\mathrm{(b)}}$ \\
      \hline
      $\sigma_{obs}$ of observation noises & \multicolumn{4}{c}{10\% of truth}  \\
      RANS mesh ($n_x \times n_y$) & \multicolumn{4}{c}{$50 \times 30$} \\ 
      number of samples $N$ & \multicolumn{4}{c}{60} \\
      \hline            
    \end{tabular} \\
  \flushleft
    (a)  see Eq.~\ref{eq:gp-kernel}, the length scale is normalized by hill crest heigth $H$.\\
    (b) $\sigma(x)$ as shown in Fig.~\ref{fig:domain_pehill}
\end{table}
 
\subsection{Prior Knowledge on Dimensionality of Uncertainty Space}
\label{sec:complex}
\begin{figure}[htbp]
  \centering
   \hspace{2em}\includegraphics[width=0.7\textwidth]{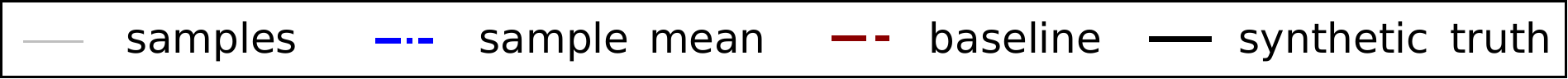}\\
   \subfloat[Prior Reynolds stress ensemble]{\includegraphics[width=0.75\textwidth]{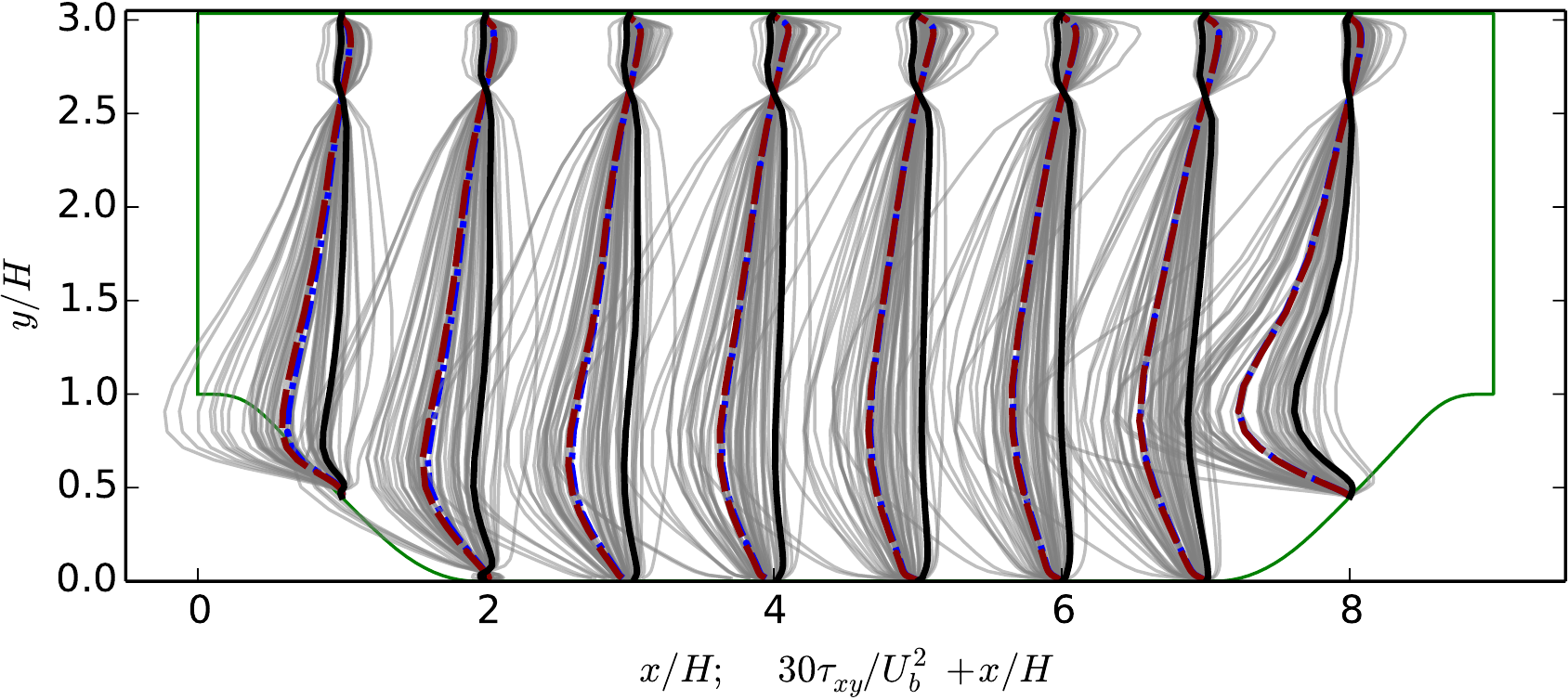}}\\
   \subfloat[Posterior Reynolds stress ensemble]{\includegraphics[width=0.75\textwidth]{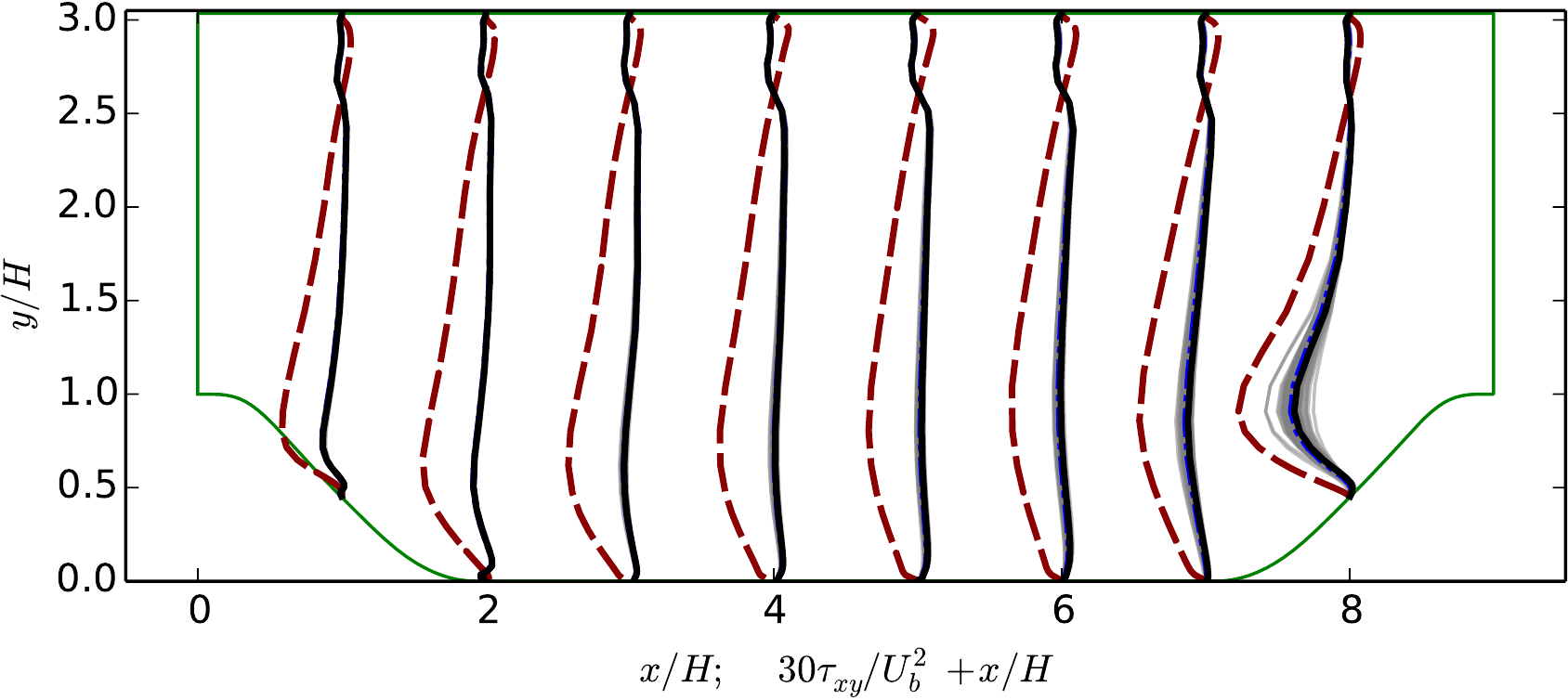}}\\
  \caption{The Prior and posterior ensembles of $\tau_{xy}$ profiles of case S1, in which the synthetic truth is used 
  and the searching space is larger than the uncertainty space where the truth resides. The ensemble profiles are shown at 
  eight locations $x/H = 1, \  \cdots, 8$, compared with synthetic truth and  baseline results. In panel (b), 
  the samples and sample mean overlap with the synthetic truth in most of the locations.}
  \label{fig:tauS1}
\end{figure}

\begin{figure}[htbp]
\centering
\hspace{2em}\includegraphics[width=0.55\textwidth]{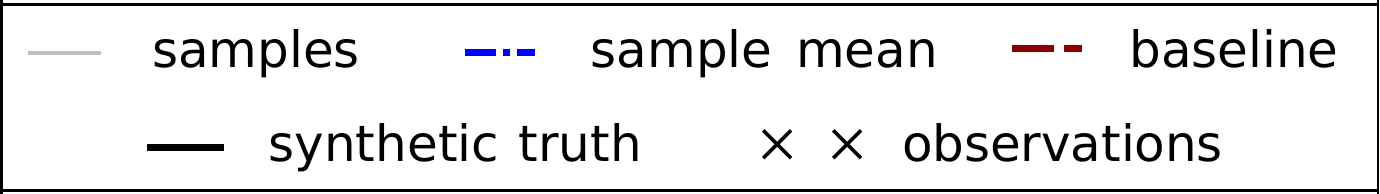}\\
\subfloat[Prior velocities ensemble]
{\includegraphics[width=0.75\textwidth]{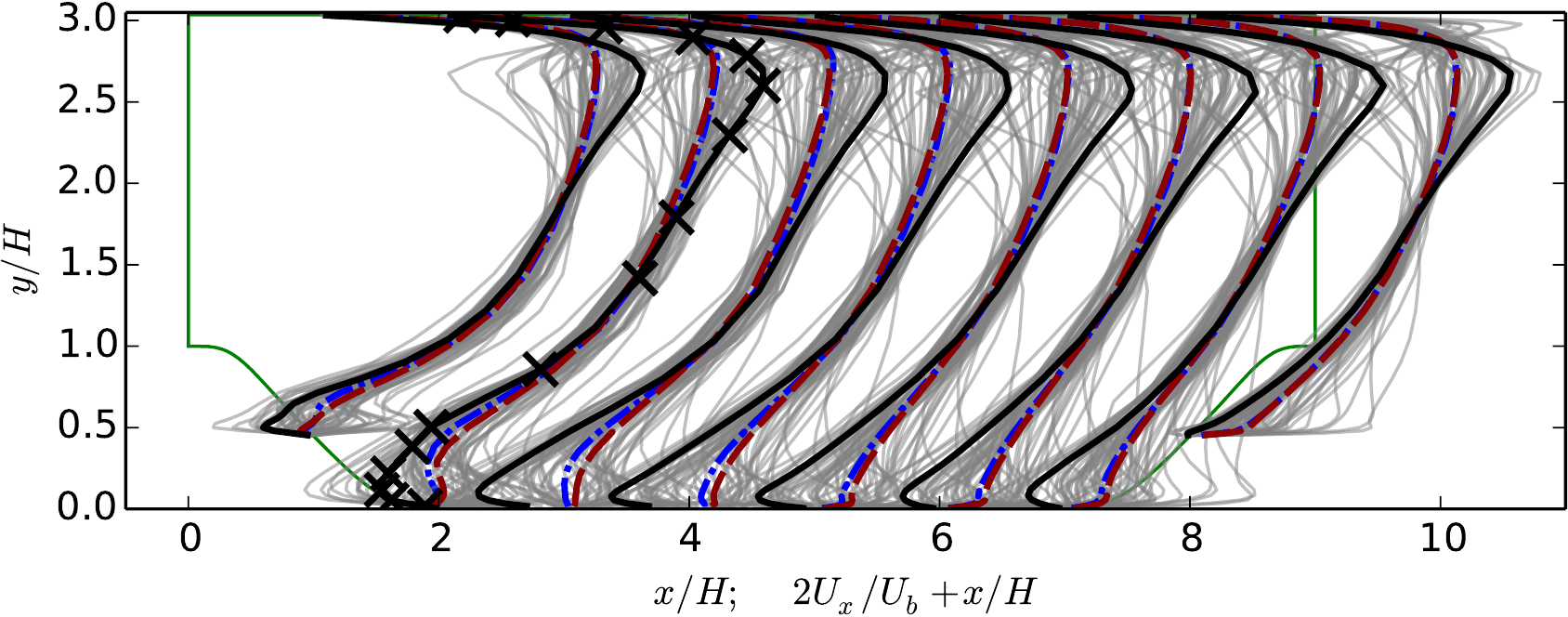}}\\
\subfloat[Posterior velocities ensemble]
{\includegraphics[width=0.75\textwidth]{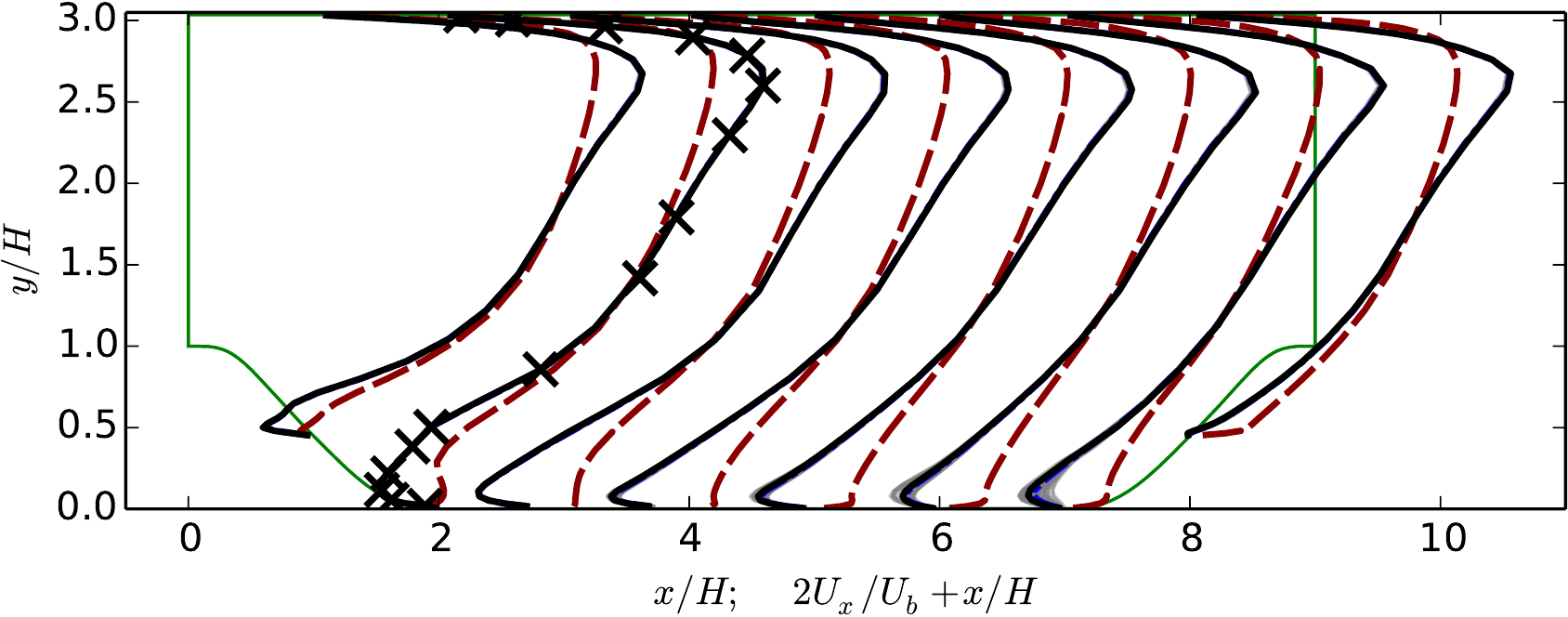}}\\
\caption{The Prior and posterior ensembles of velocity profiles of case S1.  
The locations where velocities are observed are indicated with $\times$. The ensemble profiles are shown at 
eight locations $x/H = 1, \  \cdots, 8$, compared with synthetic truth and  baseline results. In panel (b),
all the posterior samples are collapsed to the synthetic truth, and the corresponding lines are overlapped.}
\label{fig:US1}
\end{figure}

Under most circumstances, the exact dimension of the uncertainty space where the truth resides is
unknown. In order to cover the truth, we commonly span the uncertainty space with a
higher dimension. In case S1, the inversion is conducted in the uncertainties space expanded in
both the shape and amplitude of $\bs{\tau}$ (i.e., $\xi, \eta$ and $k$) with six KL modes for each field. 
Therefore, the uncertainty space has a higher dimension 
(18 modes) than that of the exact space where the 
synthetic truth resides (2 modes). The prior ensemble of Reynolds stress perturbed
in this uncertainty space is obtained, whose component $\tau_{xy}$ is shown in 
Fig.~\ref{fig:tauS1}a. It can be seen that the sample mean of $\tau_{xy}$
nearly coincides with the baseline results, and the region of ensemble 
covers the truth. This evidence indicates that the ensemble is sufficient to represent
the prior distribution. It also can be seen that the prior sample mean is biased compared
with the truth. By performing the model evaluations based on the samples of the prior 
Reynolds stresses, the prior velocity ensemble is obtained. 
Figure~\ref{fig:US1}a presents the prior ensemble of velocities $U_x$, 
which is scattered due to the large uncertainties in prior $\bs{\tau}$ ensemble.  

Synthetic observations of velocities at 16 locations along the line of $x/H = 2$
(shown in Fig.~\ref{fig:US1}) are used for inference. Figure~\ref{fig:tauS1}b 
shows the posterior $\tau_{xy}$ ensemble. We can see that its scattering 
is significantly reduced and all samples converge to the synthetic truth. 
However, some uncertainties still exist in the posterior Reynolds stress $\tau_{xy}$ and are slightly larger 
in the regions further away from the observed locations.
This indicates that the corrections on the prior ensemble become less effective, 
because when the observed and unobserved locations have a larger distance,  
the spatial correlations between the two are weak.

Figure~\ref{fig:US1}b shows the posterior velocity $U_x$ profiles, which collapse to 
the truth.   
Although uncertainties still exist in the posterior Reynolds stresses (e.g., at $x/H = 8$ in Fig.~\ref{fig:tauS1}), 
the velocities obtained based on these Reynolds stress samples are less scattered. 
The reason is likely to be that the mapping from Reynolds stress to velocity is 
not unique, and different Reynolds stress fields may map to the similar velocity fields. This issue will be further discussed in Sec.~\ref{sec:dis}

\begin{figure}[htbp]
  \centering
   \subfloat[Ensemble of $\omega$ for the $1^{\mathrm{st}}$ mode of $\delta^{\eta}$]
   {\includegraphics[width=0.5\textwidth]{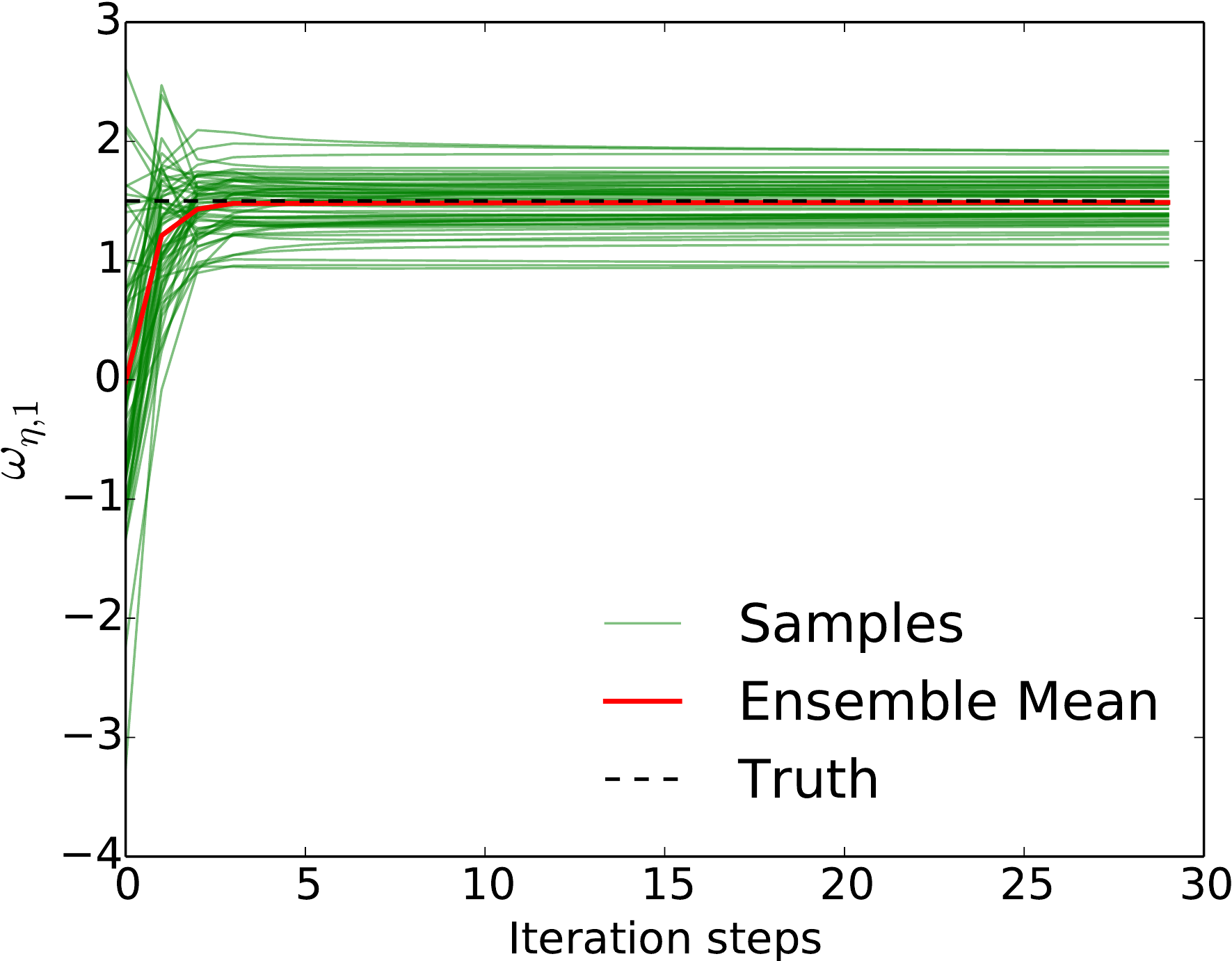}}
   \subfloat[Ensemble of $\omega$ for the $2^{\mathrm{nd}}$ mode of $\delta^{\eta}$]
   {\includegraphics[width=0.5\textwidth]{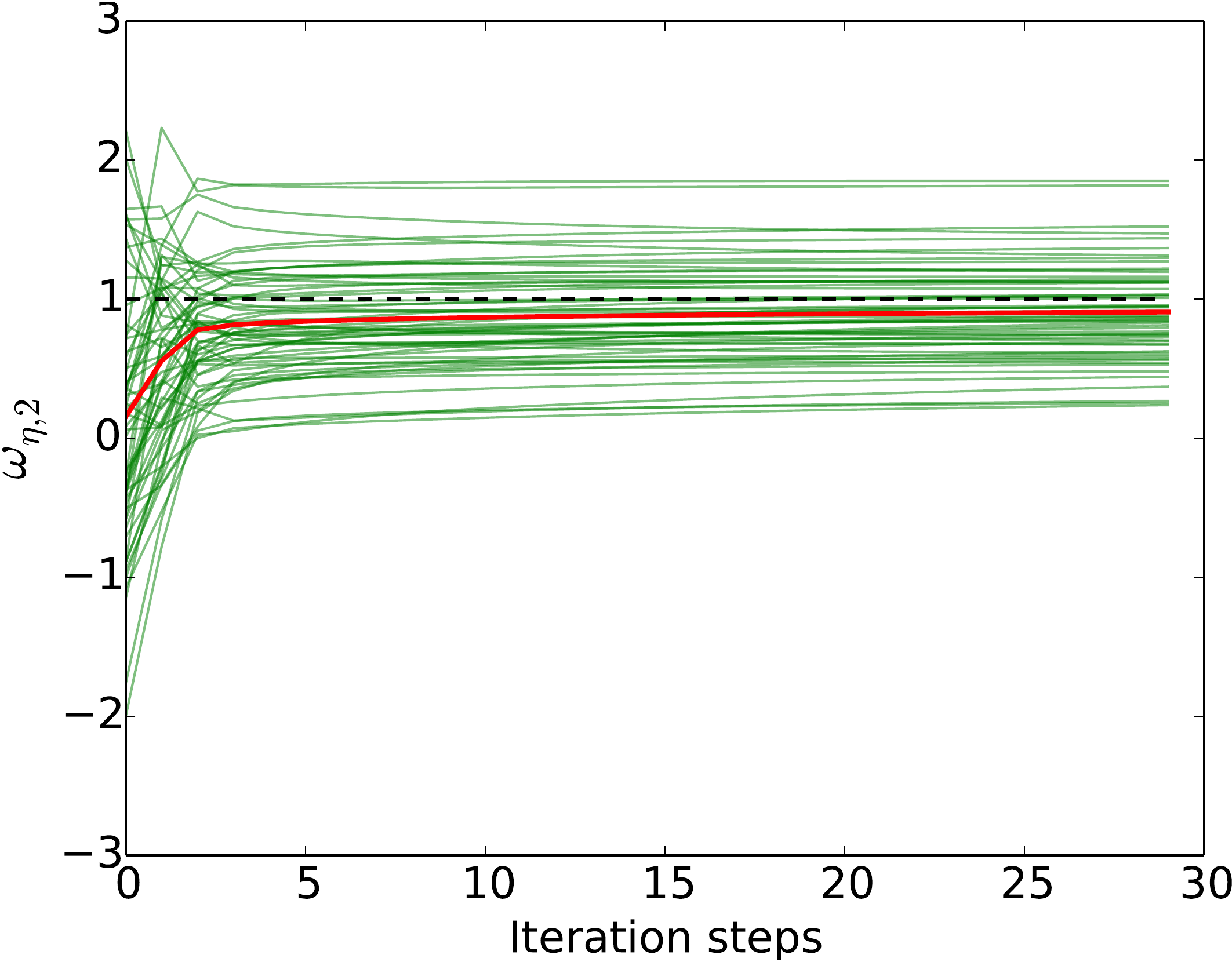}}
  \caption{Convergence histories of unknown parameters $\omega_{\eta, 1}$ and $\omega_{\eta, 2}$, which are
  the coefficients for the $1^{st}$ and $2^{nd}$ modes of $\delta^{\eta}$, respectively, for case S1. 
  The true coefficients for this synthetic case are indicated by the dashed lines.}
  \label{fig:paraHis1}
\end{figure}

The uncertainty reduction process in the proposed framework is the inversion
of the modes coefficients $\bs{\omega}$ for the Reynolds stresses discrepancy $\delta \bs{\tau}$. 
Since the posterior velocity has a good agreement with its truth, it is anticipated that 
the posterior of $\bs{\omega}$ should also converge to the truth. This anticipation can be verified in the
synthetic cases S1 and S2 where the true values of coefficients 
($\hat{\omega}_{\eta, 1} = 1.5$ and $\hat{\omega}_{\eta, 1} = 1.0$) are known. 
The convergence history of coefficients ensembles 
($\omega_{\eta, 1}$ and $\omega_{\eta, 2}$) is presented in Fig.~\ref{fig:paraHis1}. 
It shows that most samples in the $\omega_{\eta,1}$ and $\omega_{\eta,2}$ 
ensembles are initially scattered from 
$-2$ to $2$, due to the fact that they are drawn from the standard normal distribution $\mathcal{N}(0, 1)$. 
The initial mean values of $\omega_{\eta, 1}$ and $\omega_{\eta, 2}$ are
both zero, which are biased compared to their respective truths. 
Nonetheless, the ensemble mean values converge to the truths
within only a few iterations, and the scattering of the samples are slightly reduced
for both coefficients. The convergence of the inferred coefficients results in 
the successful correction of the velocity field with reduced uncertainties.   
By comparing Figs.~\ref{fig:paraHis1}a and~\ref{fig:paraHis1}b, 
we find that the converged $\omega_{\eta, 2}$
ensemble has a slightly larger variance than that of the $\omega_{\eta, 1}$ ensemble, which means that
the posterior uncertainty of $\omega_{\eta, 2}$ is slightly larger. This is because the KL modes used to 
construct the random fields are not equally weighted. The higher modes are less important than the 
lower modes. Therefore, the coefficients for the higher modes are less sensitive to
observations in the inversion process. 

\begin{figure}[htbp]
  \centering
   \subfloat[Ensemble of $\omega$ for the $1^{\mathrm{st}}$ mode of $\delta^{\eta}$]
   {\includegraphics[width=0.5\textwidth]{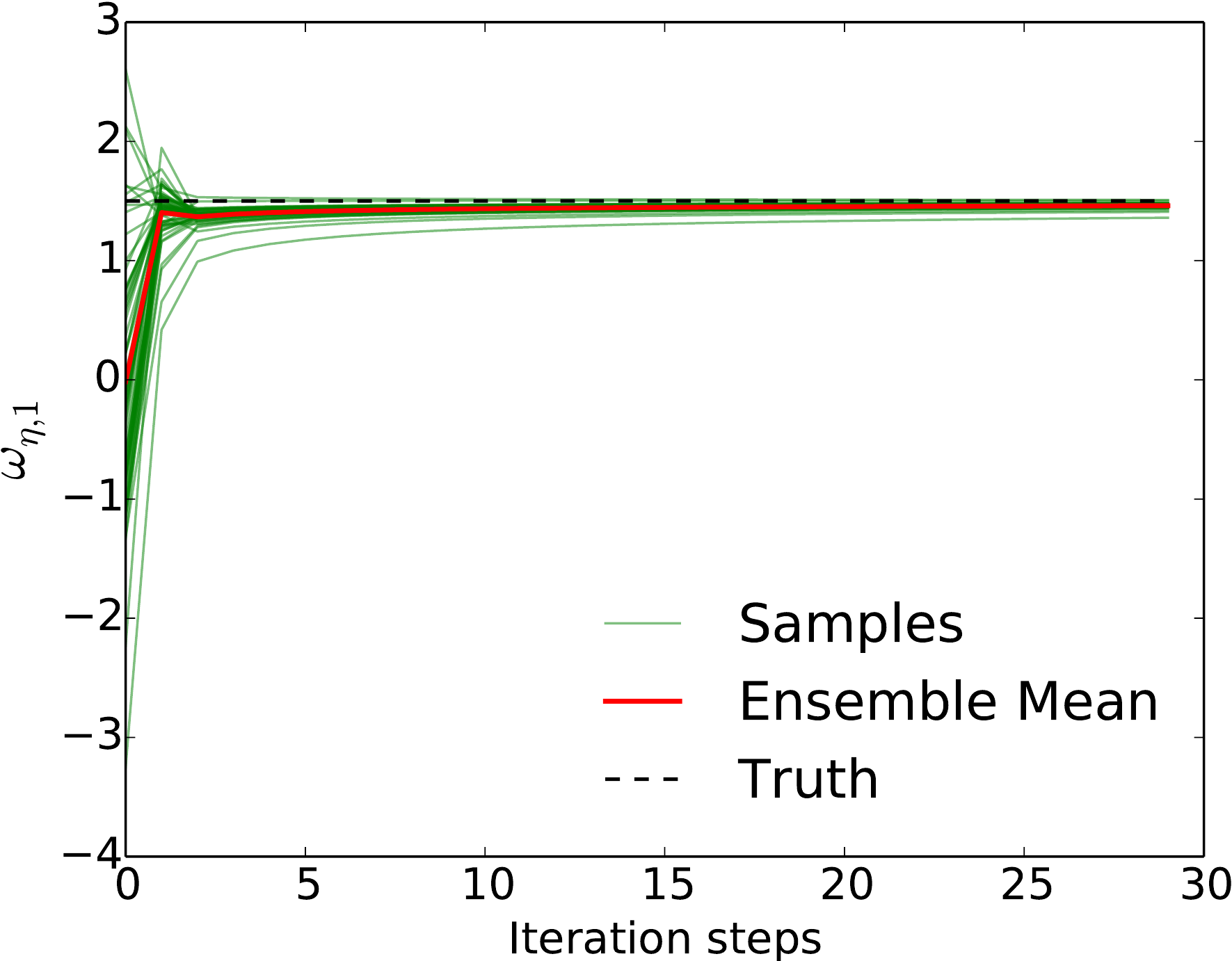}}
   \subfloat[Ensemble of $\omega$ for the $2^{\mathrm{nd}}$ mode of $\delta^{\eta}$]
   {\includegraphics[width=0.5\textwidth]{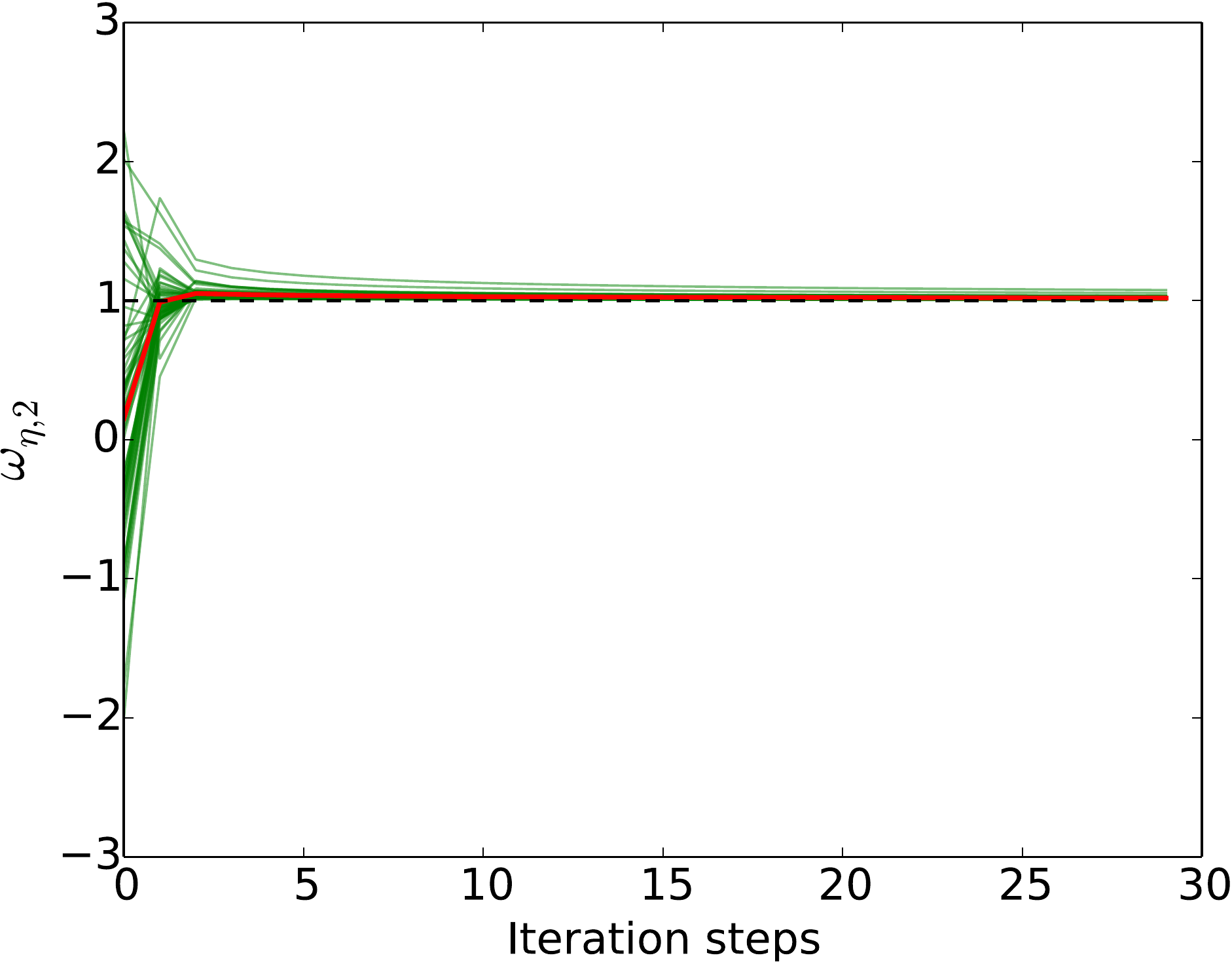}}
  \caption{Convergence histories of unknown parameters $\omega_{\eta, 1}$ and $\omega_{\eta, 2}$  
  for case S2. The searching space in this synthetic case is smaller than that of case S1. 
  The synthetic truths are indicated by the dashed lines.}
  \label{fig:paraHis2}
\end{figure}

\begin{figure}[htbp]
  \centering
  \hspace{2em}\includegraphics[width=0.7\textwidth]{./pehill-U-legend-syn-noObs}\\
   \subfloat{\includegraphics[width=0.75\textwidth]
   {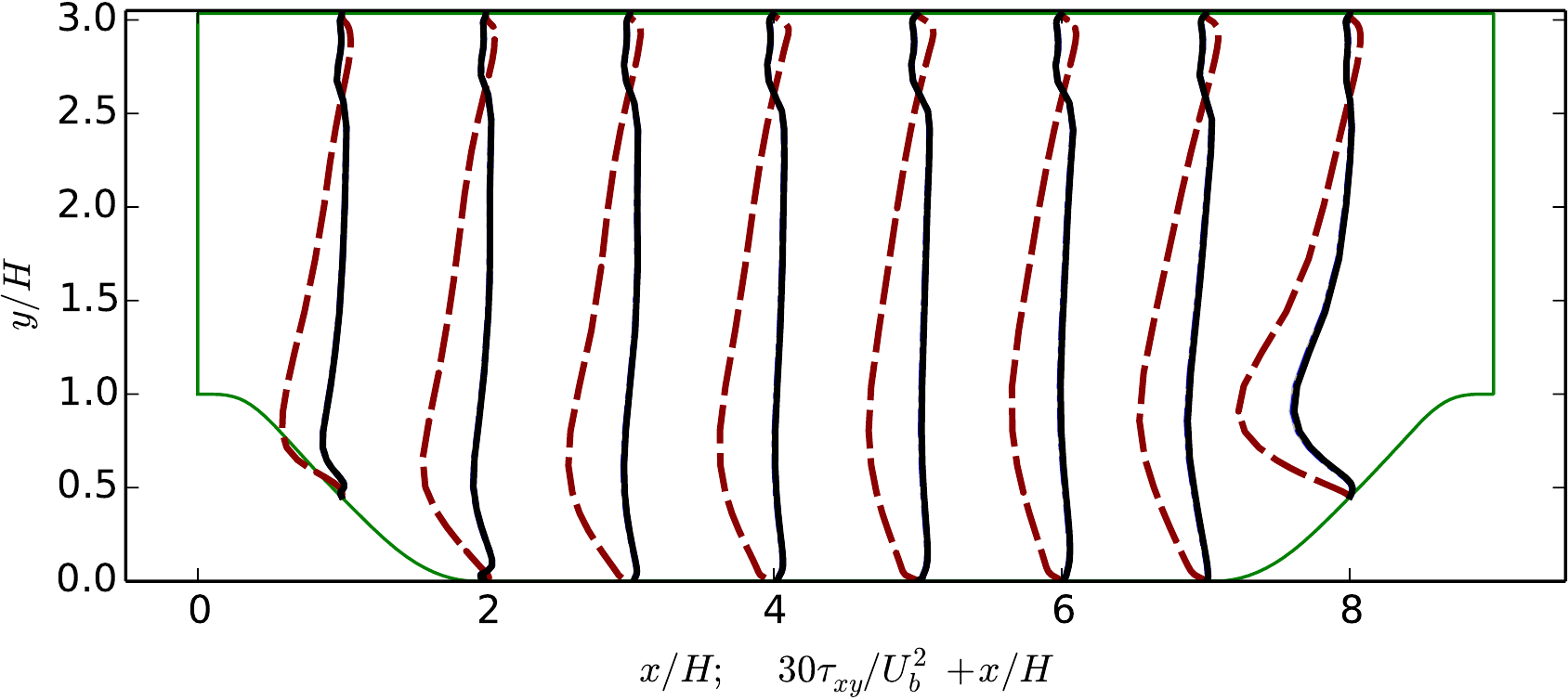}}\\
\caption{The posterior ensembles of $\tau_{xy}$ profiles of case S2, in which the synthetic truth 
is used and the searching space is the same as the uncertainty space where truth resides. The 
ensemble profiles are shown at eight locations $x/H = 1, \  \cdots, 8$, compared with synthetic truth and  baseline results.
The prior ensemble is the same as that of case S1 and is omitted for simplicity. All of the samples are collapsed to the truth, 
and thus the corresponding lines of samples and sample mean overlap with the synthetic truth.}
  \label{fig:tauS2}
\end{figure}

In case S1, uncertainties still exist in the posterior coefficients of KL modes and thus in 
the posterior Reynolds stresses ensembles. This is because the prior ensemble is sampled in an
uncertainty space with a higher dimension than that the truth resides in. 
This situation is typical in practical scenarios,
since the truth is unknown for most non-trivial problems. Therefore, the inversion is started from a
larger searching space to ensure the coverage of the truth. However, if the dimension of uncertainty space
can be reduced based on prior knowledge, the uncertainties of the inversion results can be further reduced. We demonstrate this statement with case S2, in which the inversion
is performed in a low-dimensional uncertainty space where the synthetic truth resides. In order to search this 
low-dimensional space, the prior ensemble is generated in the space spanned by two KL modes of the 
$\eta(x)$ field. All other computational parameters are
the same as those in case S1. Figure~\ref{fig:paraHis2} shows the 
convergence of the coefficients for the first and second modes of $\eta$.  
The coefficients $\omega_{\eta, 1}$ and $\omega_{\eta, 2}$ ensembles almost exactly
converge to the synthetic truths, which is in contrast to Fig.~\ref{fig:paraHis1}. 
In addition, comparison of Fig.~\ref{fig:tauS2} and Fig.~\ref{fig:tauS1}b shows that 
the uncertainties in the posterior Reynolds stress component $\tau_{xy}$ are reduced.
The posterior uncertainties are larger in the case where the searching space has higher dimension, 
especially in the regions far away from the observed location, e.g.,
along the line at $x/H = 8$. With the same amount of observation data, the inference uncertainties are reduced 
by narrowing down the dimension of searching space. The reason is that higher 
dimensional uncertainty space means more degrees of freedom that need to be inferred, which may lead to 
an ill-posed problem and pose a challenge for the inversion.

\subsection{Prior Knowledge on Variance Field $\sigma(x)$}
\label{sec:variance}
The spatial field of the perturbation variance $\sigma(x)$ 
for $\xi$, $\eta$ and $k$ reflects analyst's prior belief on the uncertainty range 
at each location. If the knowledge on it is not available a priori, we can choose
a non-informative prior with a uniform variance field $\sigma(x)$. 
However, for most flow problems, we usually have some empirical knowledge. 
For example, in the case of flow over periodic hills, there are some regions 
where RANS models are known to give poor predictions,
e.g., regions with recirculation, non-parallel free-shear flow, and the strong mean flow curvature. 
The variance field $\sigma(x)$ shown in Fig.~\ref{fig:domain_pehill} is designed to reflect 
this empirical knowledge. It can be seen that in the free-shear layer
and recirculation zone, larger perturbations of Reynolds stresses are allowed, while
the baseline RANS Reynolds stresses in other regions are assumed to be relatively
reliable.

\begin{figure}[!htbp]
  \centering
  \hspace{2em}\includegraphics[width=0.55\textwidth]{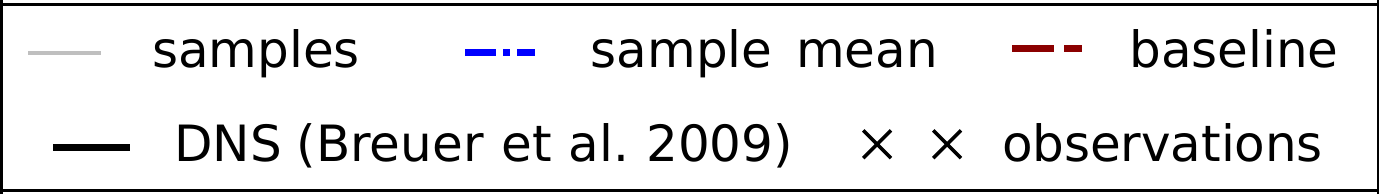}\\
   \subfloat[$U_x$ with non-informative $\sigma(x)$]
   {\includegraphics[width=0.75\textwidth]{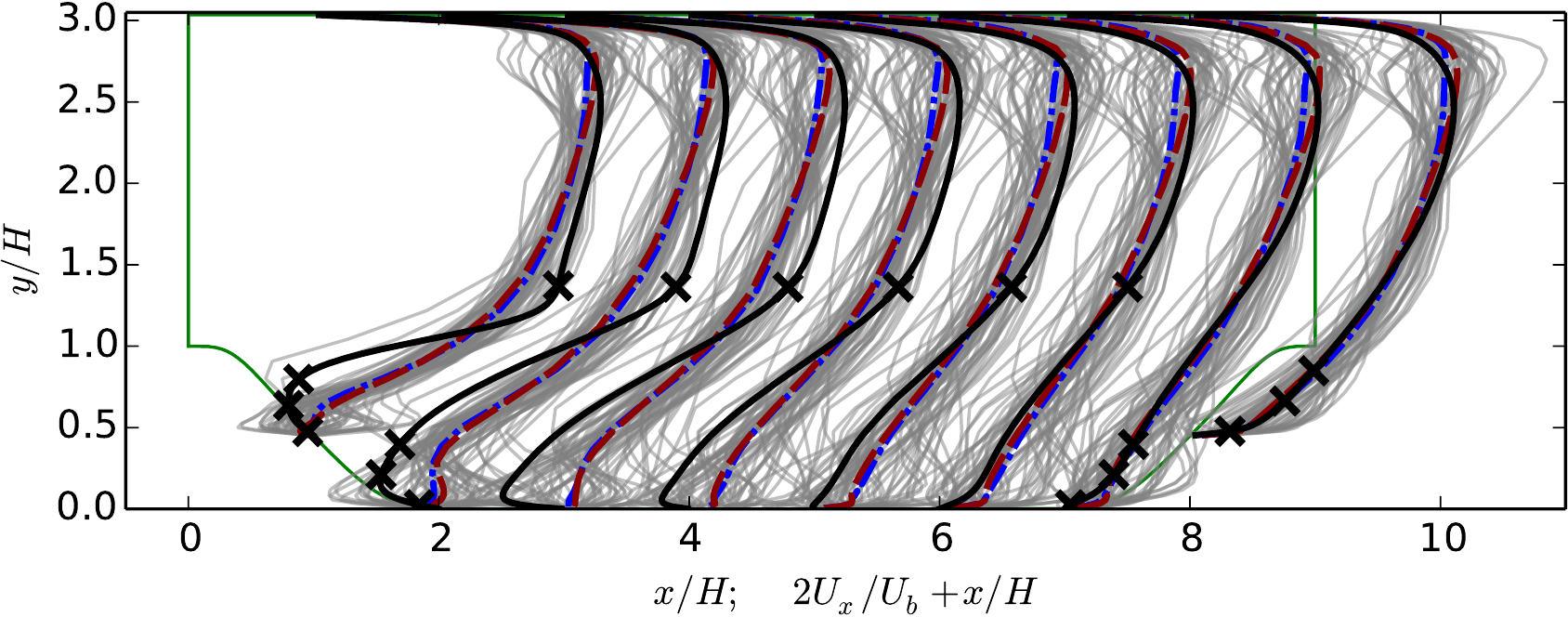}}\\
   \subfloat[$U_x$ with informative $\sigma(x)$]
   {\includegraphics[width=0.75\textwidth]{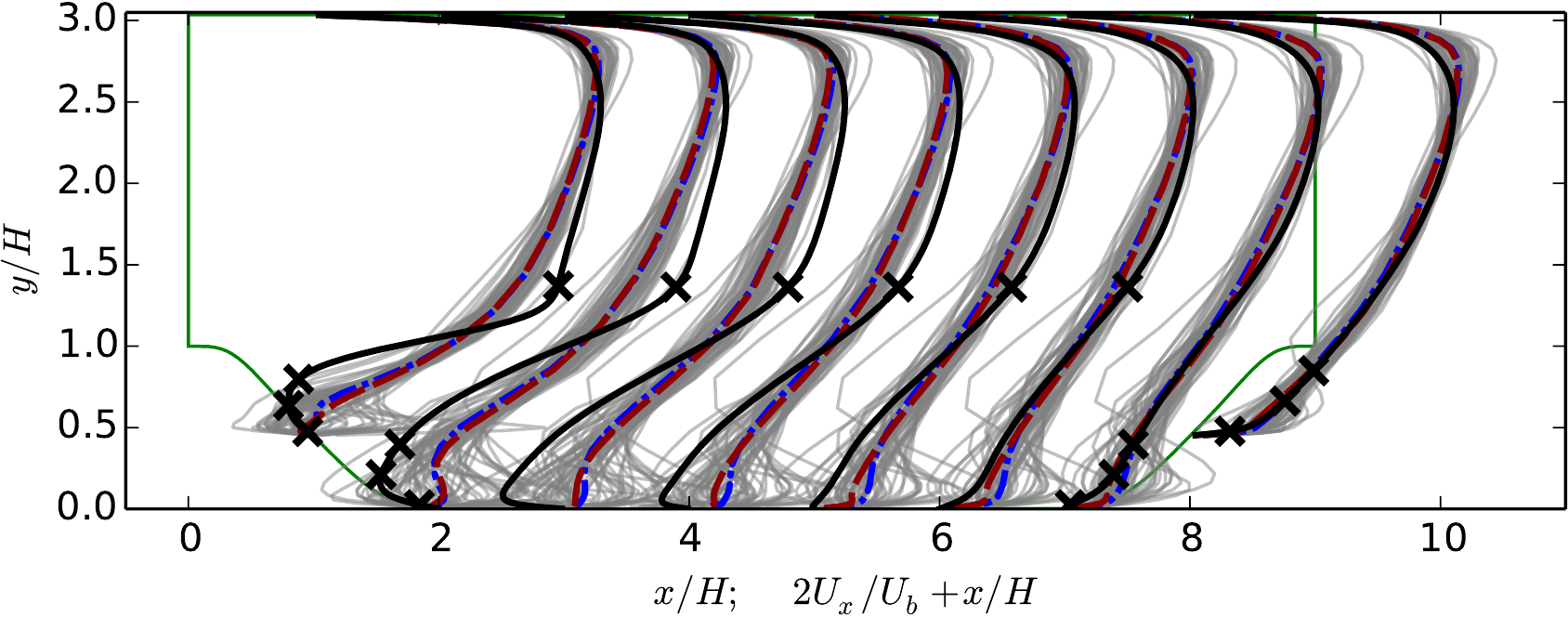}}\\
  \caption{
  The prior ensembles of velocity $U_x$ profiles of case D1 and case D2, in which the observations are from DNS benchmark.   
  (a) The variance field $\sigma(x)$ of case D1 is spatially uniform ($\sigma(x)=0.7$),
  (b) The variance field $\sigma(x)$ of case D2 is informative ($\sigma_{min}=0.2$, $\sigma_{max}=0.7$).
  The ensemble profiles are shown at eight locations $x/H = 1, \  \cdots, 8$, 
  compared with the baseline results and DNS benchmark.
  The locations where velocities are observed are indicated with $\times$. 
 }
  \label{fig:U_pehillPrior}
\end{figure}

\begin{figure}[!htbp]
  \centering
  \hspace{2em}\includegraphics[width=0.55\textwidth]{./pehill-U-legend-dns}\\
   \subfloat[$U_x$ with non-informative $\sigma(x)$]
   {\includegraphics[width=0.75\textwidth]{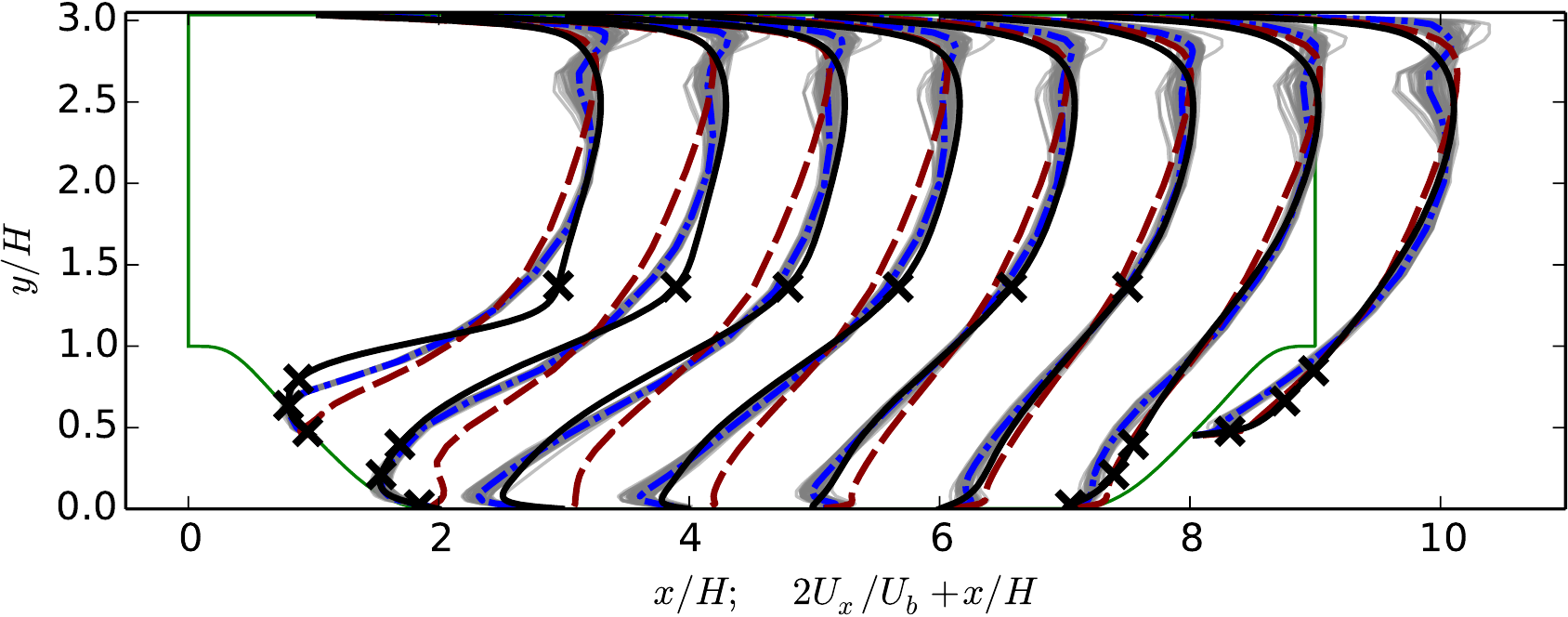}}\\
   \subfloat[$U_x$ with informative $\sigma(x)$]
   {\includegraphics[width=0.75\textwidth]{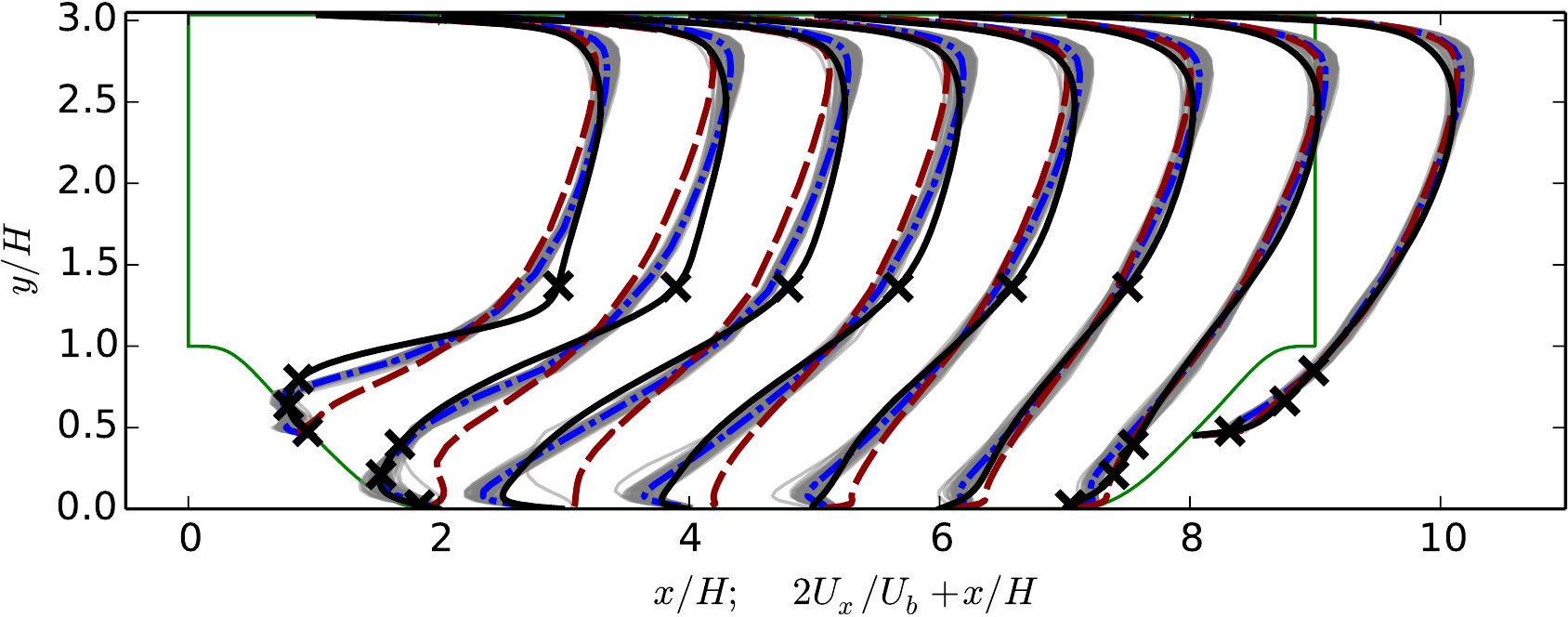}}\\
  \caption{
  The posterior ensembles of velocity $U_x$ profiles of case D1 and case D2, in which the observations are from DNS benchmark.   
  (a) The variance field $\sigma(x)$ of case D1 is spatially uniform ($\sigma(x)=0.7$),
  (b) The variance field $\sigma(x)$ of case D2 is informative ($\sigma_{min}=0.2$, $\sigma_{max}=0.7$).
  The ensemble profiles are shown at eight locations $x/H = 1, \  \cdots, 8$, 
  compared with the baseline results and DNS benchmark.
  The locations where velocities are observed are indicated with $\times$. }
  \label{fig:U_pehillPost}
\end{figure}

The prior of velocity profiles obtained based on non-informative and informative variance fields 
are shown in Figs.~\ref{fig:U_pehillPrior}a and \ref{fig:U_pehillPrior}b, respectively. 
It can be seen that the velocity samples are appreciably scattered in the entire domain
if uniform $\sigma(x)$ is employed. 
Such a large scattering is attributed to the lack of physical prior knowledge.
In contrast, the scattering of the prior velocity profiles is constrained by incorporating 
the physical knowledge on the variance field.  This is shown in Fig.~\ref{fig:U_pehillPrior}b. 
Therefore, the prior with uniform $\sigma(x)$ field has an uncertainty space with spatially equal variance,
while the uncertainties are reduced in the regions where RANS predictions are more accurate 
based on the informative $\sigma(x)$ field. The merits of incorporating empirical 
knowledge can be clearly demonstrated by comparing the posterior velocity profiles of cases D1 and
D2, which are shown in Figs.~\ref{fig:U_pehillPost}a and~\ref{fig:U_pehillPost}b, respectively.
Compared to the prior velocity profiles, both posterior results are improved since all of the velocities 
samples converge to the DNS benchmark, and their scattering is largely reduced.
This indicates that the proposed framework improves the model predictions (especially in the recirculation regions) 
even with a non-informative variance field, i.e., a constant $\sigma(x)$.    
However, in Fig.~\ref{fig:U_pehillPost}a, the scattering of posterior velocity samples obtained with 
the uniform $\sigma(x)$ field is still larger than that with informative $\sigma(x)$ field in
case S2 (shown in Fig.~\ref{fig:U_pehillPost}b). 
Especially near the top of the domain, the uncertainty in velocity ensemble is even larger and 
the sample mean velocity is slightly distorted compared to the truth and is unphysical.   
The reason is that the $\sigma(x)$ field introduces relatively large perturbation in this region,
where no observation is available nearby to constrain the uncertainties in the posterior ensemble. 
In contrast, the results obtained with an informative prior shown in Fig.~\ref{fig:U_pehillPost}b has
a much better agreement with the truth in the
region near the upper wall. It shows that the posterior distribution can be improved by 
adopting an informative variance field with prior knowledge, as it reduces unnecessary perturbations. 

\begin{figure}[!htbp]
  \centering
  \hspace{2em}\includegraphics[width=0.7\textwidth]{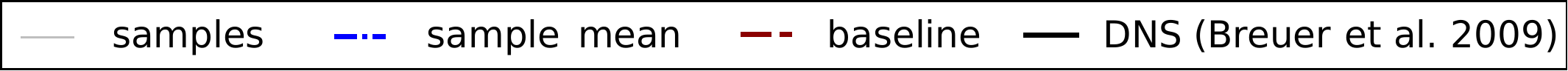}\\
   \subfloat[Wall shear stress of case D1]{\includegraphics[width=0.45\textwidth]{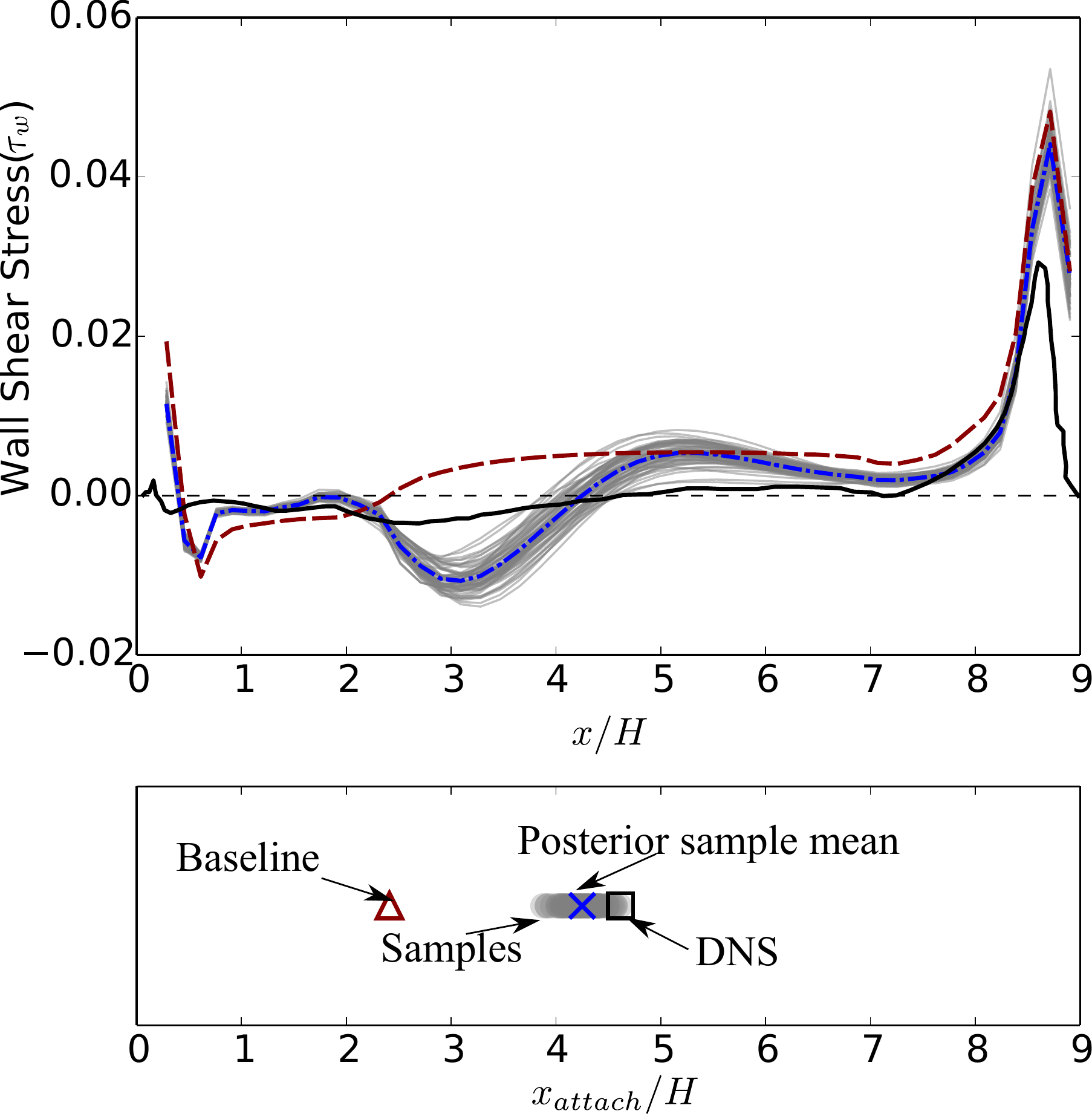}}\hspace{0.5em}
   \subfloat[Wall shear stress of case D2]{\includegraphics[width=0.45\textwidth]{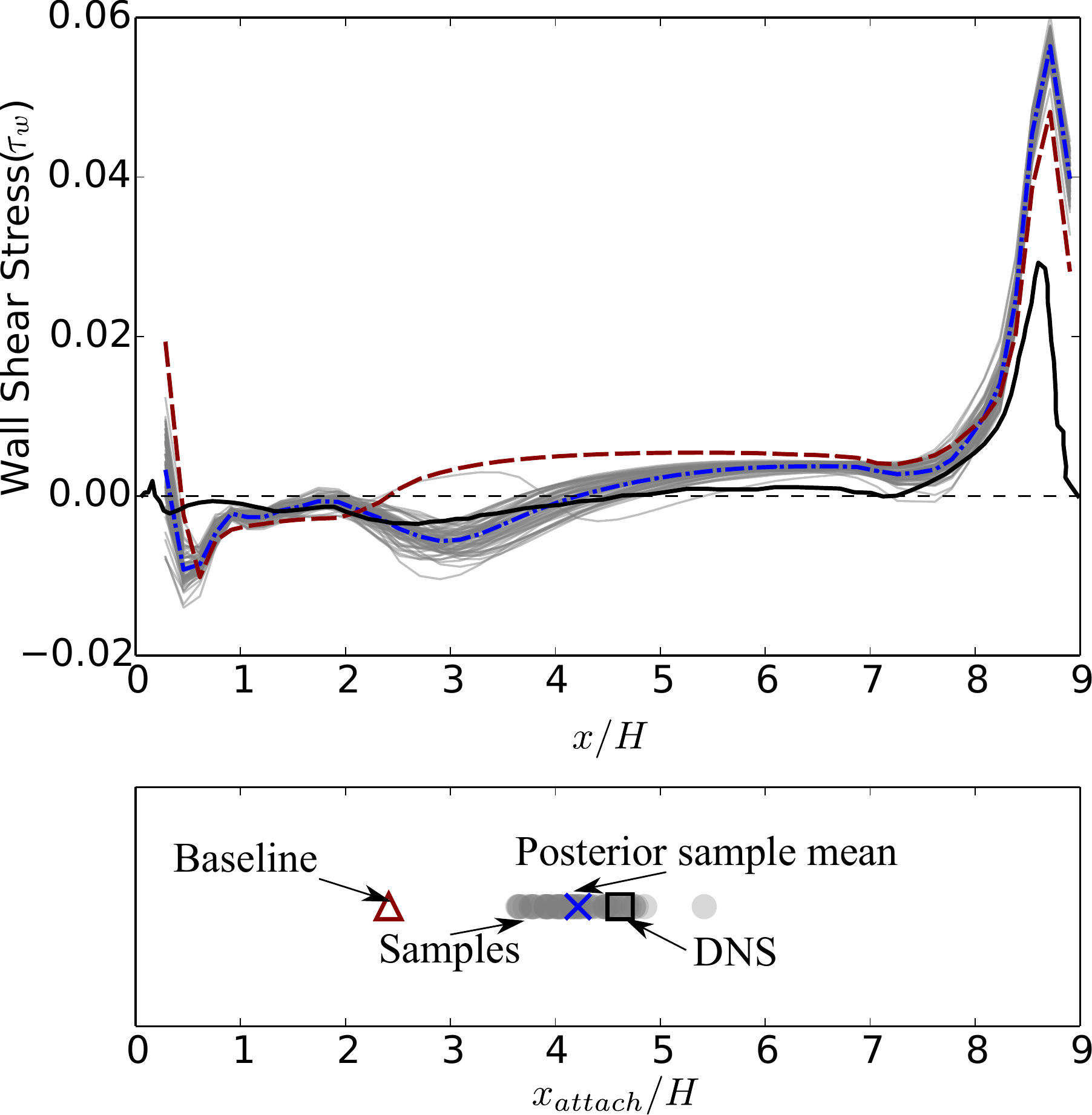}}\\
   \subfloat[Reattachment point of case D1]{\includegraphics[width=0.45\textwidth]{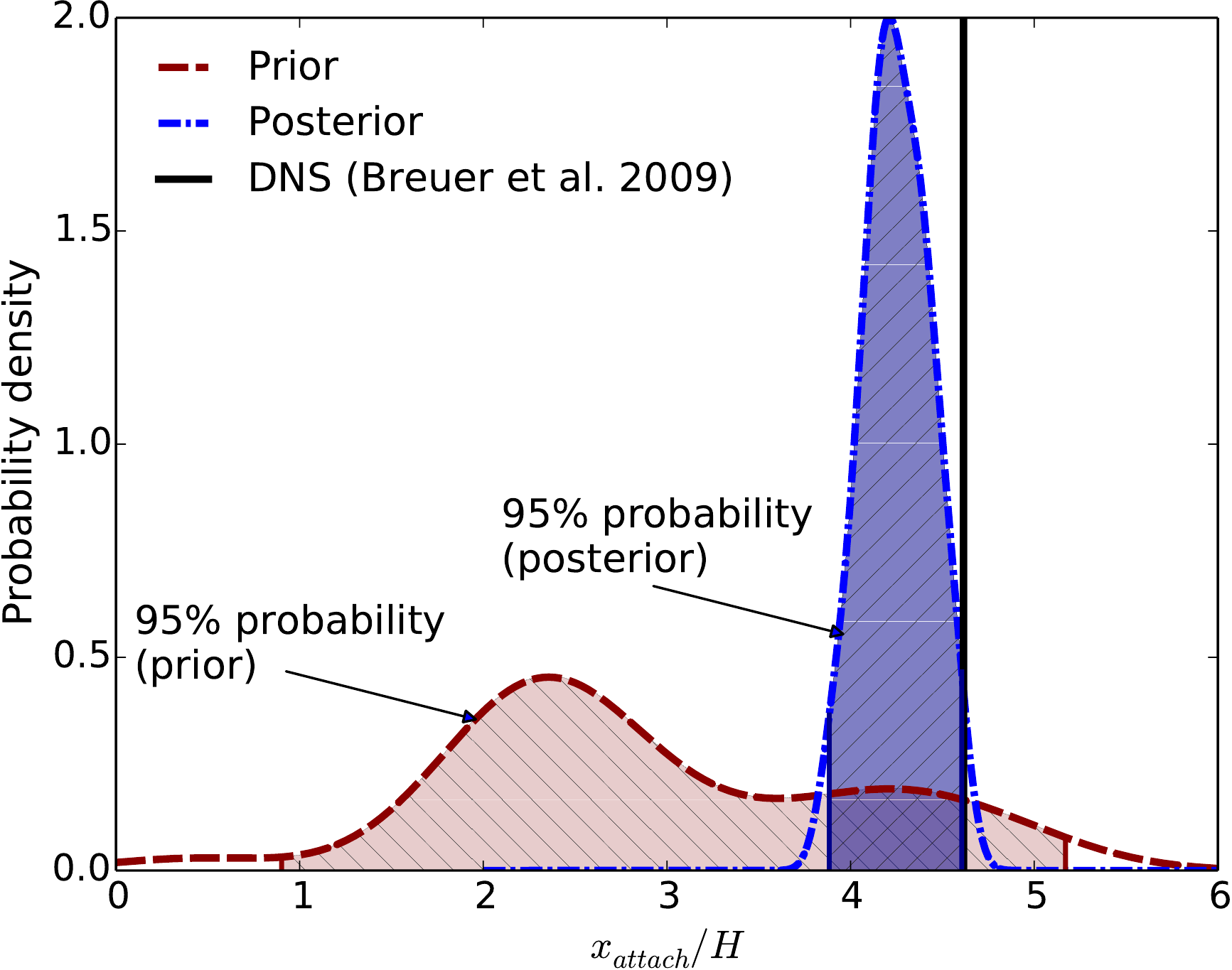}}\hspace{0.5em}
   \subfloat[Reattachment point of case D2]{\includegraphics[width=0.45\textwidth]{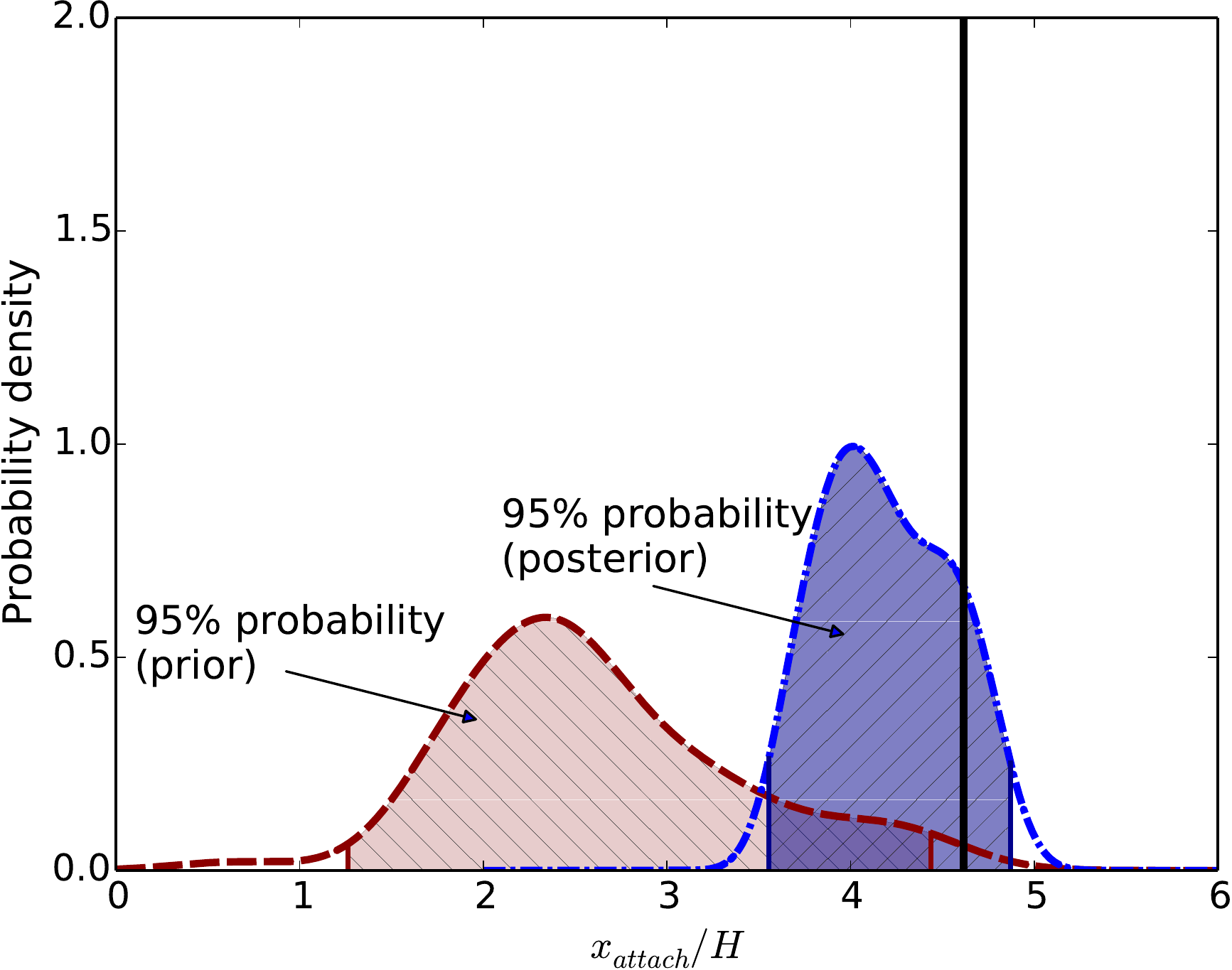}}\\
  \caption{
  The posterior wall shear stress and reattachment point obtained with uniform, non-informative $\sigma$ field (case D1) 
  and from informative $\sigma$ field (case D2). The change of wall shear stress from negative to positive 
  indicates the end of the recirculation zone. (a) Wall shear stress of case D1;
   (b) Wall shear stress of case D2; (c) Reattachment point 
   of case D1; (d) Reattachment point of case D2.}
  \label{fig:TauW_pehill}
\end{figure}

Figures~\ref{fig:TauW_pehill}a and~\ref{fig:TauW_pehill}b show the posteriors of the other two QoIs, 
wall shear stress $\tau_w$ and reattachment point $x_{attach}$, for cases
D1 and D2, respectively. We can see that both $\tau_w$ and $x_{attach}$ obtained
from baseline RANS simulation deviate from the DNS benchmark significantly. The RANS predicted
recirculation region is much smaller than the truth. The posterior means of reattachment point 
$x_{attach}$ of both cases
D1 and D2 have better agreements with the truth. However, the posterior
distribution of $x_{attach}$ obtained with the non-informative $\sigma(x)$ is overconfident with a bias, 
which can be clearly seen in the comparison of probability density 
functions (PDF) as shown in Fig.~\ref{fig:TauW_pehill}c. 
The $x_{attach}$ distribution barely covers the truth and the mean is biased. 
Specifically, it can be seen in Fig.~\ref{fig:TauW_pehill}a that $\tau_w$ is overcorrected from
$x/H = 2.5$ to $x/H = 4$ and undercorrected from $x/H = 4.5$ to $x/H = 7.5$. 
Significant improvement is achieved when an informative $\sigma(x)$ is adopted, which is shown
in Figs.~\ref{fig:TauW_pehill}b and~\ref{fig:TauW_pehill}d. We can see that in most part of the region 
(between $x/H = 1$ and $8$), the posterior ensemble has a better agreement with the benchmark.
Moreover, the posterior PDF is wider than that of case D2 and well covers the truth, indicating that
the overconfidence existed in Fig.~\ref{fig:TauW_pehill}c is reduced. All of these results demonstrate 
the merits of incorporating empirical knowledge into the $\sigma(x)$ field. It is noted 
that in the vicinity of the hill crest (i.e., near $x/H = 0.5$ and $x/H =8.5$), the posterior ensembles of both 
cases show less improvement. This is because the flow in the region with rapid 
spatial variations has a relative small length scale and thus a weak correlation with the flow in other regions. 
Consequently, the corrections are not effective due to the weak correlations between this region and the regions with 
observations. 

\subsection{Prior Knowledge on Experimental Design}
\label{sec:experi}

\begin{figure}[!htbp]
  \centering
  \hspace{2em}\includegraphics[width=0.55\textwidth]{./pehill-U-legend-dns}\\
   \subfloat[$U_x$ with non-informative experimental design]{\includegraphics[width=0.75\textwidth]{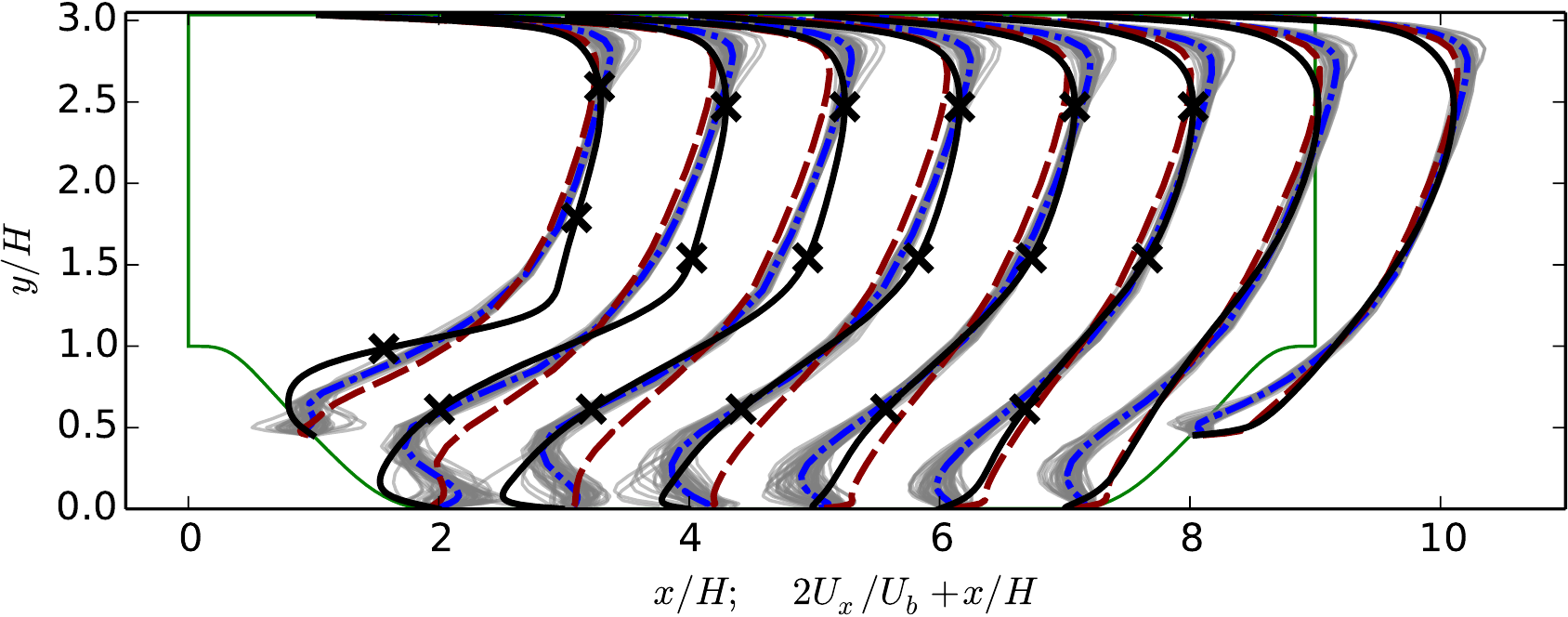}}\\
   \subfloat[$U_x$ with informative experimental design]{\includegraphics[width=0.75\textwidth]{./Ux-pehill-posterior}}\\
  \caption{
  The posterior velocity $U_x$ profiles of case D2. (a) The observations are uniformly  
  arranged across the domain; (b) The observations are arranged based on the prior 
  knowledge of the flow. The total numbers of observations in (a) and (b) are the same. }
  \label{fig:U_obs_pehill}
\end{figure}

In addition to the amount of observation data, the arrangement of observed locations 
is also a crucial factor, as it affects the inversion results of model discrepancy and 
corresponding corrections to the predicted flow field. Since the model discrepancies vary
location by location, the observed information is weighted. 
Designing the layout of the measuring points to achieve a more effective use of observation data
is referred to as ``experimental design''. In this section we will
demonstrate that prior knowledge will also improve the experimental design 
for quantifying and reducing the model-form uncertainties. 
Two principles for experimental design should be mentioned in the
proposed framework. First, the observations should be placed in the regions where the RANS
predictions are relatively unreliable (e.g., the recirculation area). Second, more observations should be allocated to 
the regions where the length scale of flow is small. This is because the inference 
and correction of model discrepancies in the unobserved regions are based on their correlations with
the observed locations.
To demonstrate the effects of prior knowledge in the experimental design, 
another scenario of case D2 with a different arrangement of observations is investigated. 
Instead of placing the observations with informative prior based on two principles mentioned above, we
uniformly arrange the same number of observations used in the case D2 in
the entire flow field, which are shown in Fig.~\ref{fig:U_obs_pehill}a.  

The posterior velocity profiles of case D2 with non-informative and informative 
experimental designs are compared in Figs.~\ref{fig:U_obs_pehill}a and~\ref{fig:U_obs_pehill}b,
respectively. Figure~\ref{fig:U_obs_pehill}a shows that the bias in the recirculation zone is 
only partially corrected  with a uniform layout of observations. 
The reverse flow near the wall ($y/H < 0.5$) is still poorly predicted, since no observation is
available in the nearby region. Moreover, overcorrection can be seen in the downstream
($x/H > 5$), where the baseline RANS prediction is relatively accurate. The reason is that 
the length scale of flow in the near wall region is small, while no observation is placed
in the windward side of hills ($x/H = 7$ to $8$).  Consequently,
the spatial correlation is not strong enough to correct the profiles and constrain the uncertainties 
by incorporating the information from the observations. 

\section{Discussion}
\label{sec:dis}

\subsection{Effect of Observation Data Versus Effect of Empirical Prior Knowledge}

\begin{figure}[htbp]
  \centering
  \hspace{2em}\includegraphics[width=0.7\textwidth]{./pehill-U-legend-syn-noObs}\\
   \subfloat{\includegraphics[width=0.75\textwidth]
   {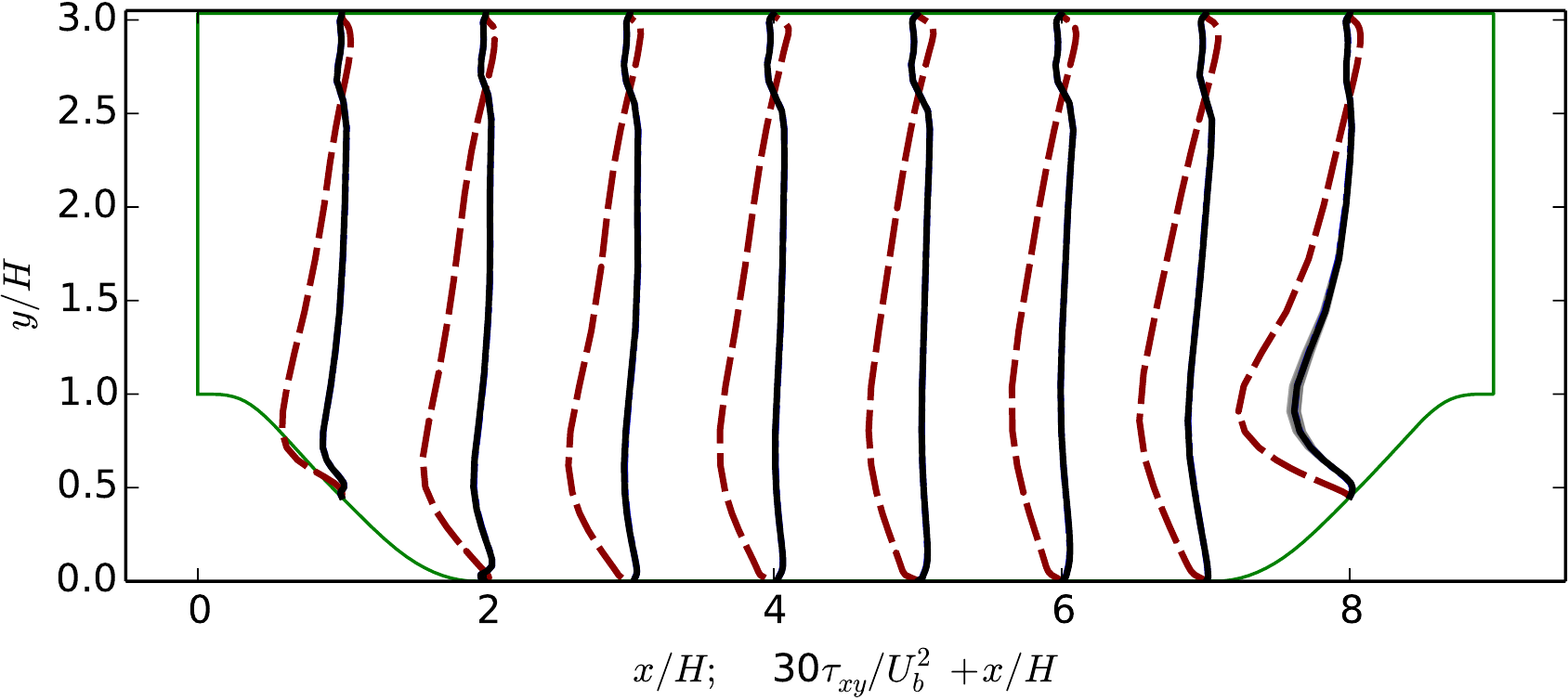}}\\
\caption{The posterior ensembles of $\tau_{xy}$ profiles of case S1 with more velocity observations 
(32 compared to 16 in the original case S1). The ensemble profiles are shown
 at eight locations $x/H = 1, \  \cdots, 8$, compared with synthetic truth and  baseline results. 
 All of the samples are collapsed to the truth, 
 and thus the corresponding lines of samples and sample mean overlap with the synthetic truth.}
  \label{fig:tauSynObs}
\end{figure}

In Bayesian frameworks, uncertainties in the posterior distribution can be reduced by incorporating 
observation data. This statement can be confirmed by exploring another scenario of 
case S1 in which the velocities at additional 16 points at $x/H = 7$ are observed. 
Figure~\ref{fig:tauSynObs} shows the inferred posterior ensemble of Reynolds stress component $\tau_{xy}$ 
with more observation data. By comparing Figs.~\ref{fig:tauS1}b and~\ref{fig:tauSynObs}, 
we can see that the scattering of the samples is significantly reduced with more observation data,
especially near the regions where the additional observation points are introduced. Therefore, incorporating more
observation data improves the inferred quantities (i.e., Reynolds stress) and reduces the corresponding
uncertainties.

However, for most engineering problems the amount of data is limited and is
insufficient to drive a physics-neutral, data-based
statistical frameworks proposed in the statistical community. On the other hand, empirical knowledge
 is available for many practical applications thanks to the accumulated experiences in the 
 engineering community. We claim that an important advantage of the proposed framework is to provide a rigorous 
 Bayesian approach incorporating various sources of empirical knowledge. 
 The incorporated empirical knowledge complements the inadequate observation data 
 and thus further reduces the model-form uncertainties in RANS simulations.
  
\subsection{Non-uniqueness of Mapping from Inferred Quantities to Observed Quantities}
An essential component of the proposed model-form uncertainty quantification and reduction framework is the
inversion of discrepancies in the Reynolds stress tensor. However, the inferred quantities (i.e., discrepancy in the
Reynolds stress field) do not uniquely map to the corresponding observed quantities (i.e., velocities). 
This is due to the ill-possedness of the problem. That is, the mapping from velocity field to Reynolds stress field
has multiple solutions. Such ill-possedness is common in many inversion problems \cite{o1986statistical}.   

\begin{figure}[htbp]
\centering
\hspace{2em}\includegraphics[width=0.7\textwidth]{./pehill-U-legend-syn-noObs}\\
\subfloat{\includegraphics[width=0.8\textwidth]{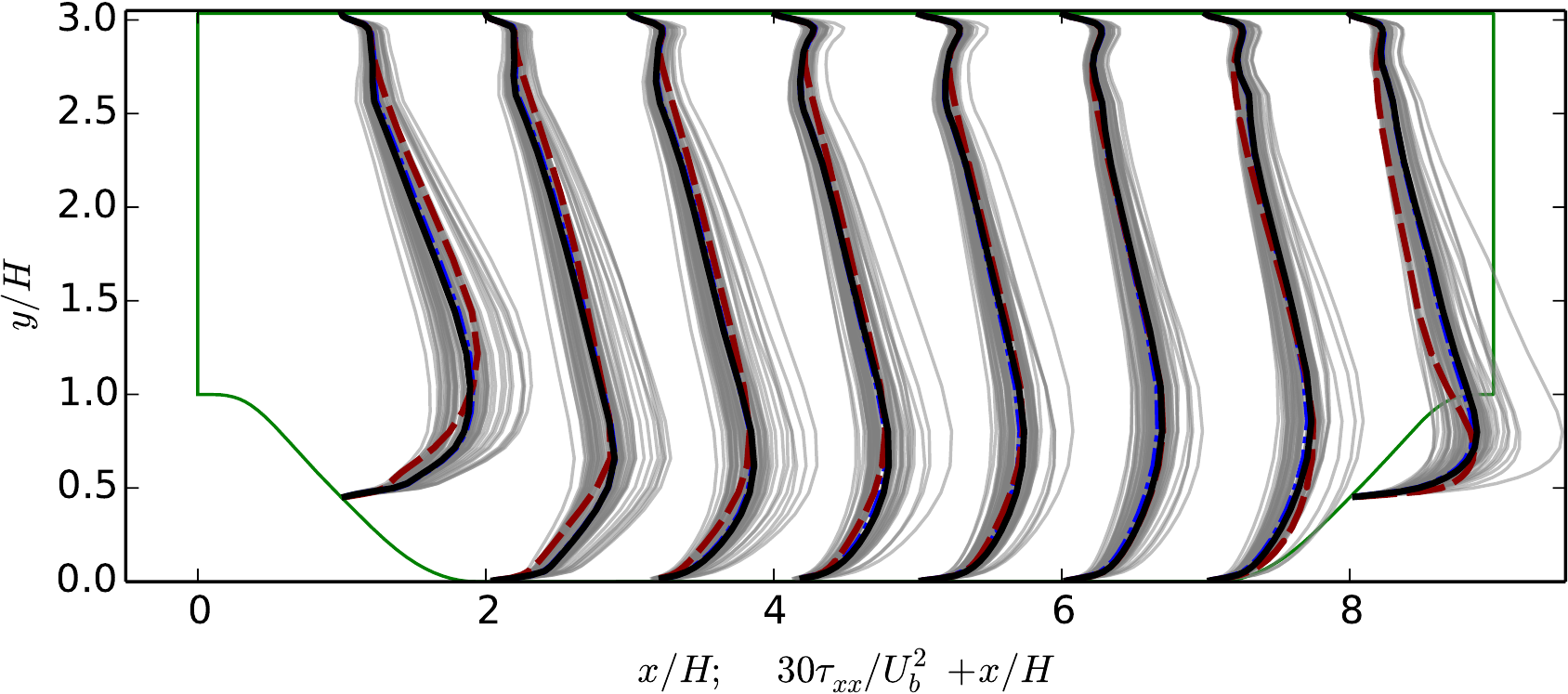}}\\
\caption{The posterior ensembles of $\tau_{xx}$ profiles of case S1. The ensemble profiles 
are shown at eight locations $x/H = 1, \  \cdots, 8$ of case S1, compared with synthetic truth and  baseline results.}
\label{fig:tauxxSyn1}
\end{figure}

The non-unique mapping can be demonstrated by the case S1. 
As shown in Fig.~\ref{fig:US1}, the posterior velocity profiles
converge to the truth and the uncertainties (i.e., scattering of velocity samples) are nearly eliminated 
based on the synthetic observation data. However, the inferred ensemble of modes coefficients for Reynolds stress discrepancy
are still scattered (shown in Fig.~\ref{fig:paraHis1}). This evidence implies that the Reynolds stress fields are 
still scattered, even though the corresponding velocity fields are almost the same. 
The scattering of Reynolds stress fields also can be seen clearly in posterior ensemble of $\tau_{xx}$, shown in Fig.~\ref{fig:tauxxSyn1}. 
The scattering range of posterior $\tau_{xx}$ ensemble is large in the entire domain, which indicates the velocities are 
not sensitive to $\tau_{xx}$ in this problem. 
Nonetheless, based on the concept of ``goal-oriented uncertainty quantification''~\cite{duraisamy2012goal}, the merits of proposed 
framework can be justified depending on the quantities of interest (QoI). 
If QoI is the velocity field or wall shear stress, the predicted results are significantly improved through
the framework. On the other hand, if the Reynolds stress is the QoI, more prior knowledge or observation data are needed 
to constrain the problem of ill-posedness. It has been shown in case S1 that the true Reynolds stress can also be inferred exactly, 
so long as sufficient data and prior knowledge are given.

\section{Conclusion}
\label{sec:con}
In this paper, we discussed the merits of incorporating various prior knowledge into the 
proposed Bayesian framework for model-form uncertainty quantification and reduction in RANS simulations.
The prior knowledge considered in the framework includes the smoothness and realizability of Reynolds stress tensor field, 
overall understanding on the coherent structures of the flow, and the empirical experiences of RANS modeling. 
We examined the merits of incorporating the prior knowledge on the dimension of uncertainty space, spatial field of variance
and experimental design by using four test cases. The simulation results demonstrate that 
the posterior QoI predictions can be significantly improved by incorporating physical prior knowledge into the Bayesian inference.  
Incorporating the prior information enables a more efficient usage of limited observation data. Meanwhile, the study
also suggests that the proposed framework provides a relatively rigorous way to express and incorporate the 
existing empirical knowledge on RANS modeling. Based on the proposed framework, the empirical knowledge 
accumulated in decades of engineering practices can be used to improve the quantification and reduction of model-form uncertainties
in RANS simulations. 

\appendix
\section{Notation}

\begin{tabbing}
  0000000\= this is definition\kill 
  $k$ \> turbulent kinetic energy \\
  $l$ \> length scale of Gaussian random field \\
  $m$ \> number of modes for each parameter\\
  $x$ \> spatial coordinate   \\
  $C$ \> Barycentric coordinate \\
  $K$ \> Gaussian kernel\\
  $H$ \> crest height\\
  $N$ \> number of samples\\
  $Re_b$ \> Reynolds stress\\
  $U_b$ \> bulk velocity\\
  $U_x$ \> velocity component in $x$ \\
  $\mathbf{a}$ \> anisotropy tensor\\
  $\mathbf{I}$ \> second order identity tensor \\
  $\mathbf{v}$ \> orthonormal eigenvector component  of $\mathbf{a}$\\  
  $\mathbf{V}$ \> orthonormal eigenvectors of $\mathbf{a}$\\
  $\mathcal{GP}$ \> Gaussian random field  \\
  
{Greek letters}{}\\
  $\bs{\tau}$ \> Reynolds stress \\
  $\tau$ \> component of Reynolds stress \\
  $\lambda$ \> eigenvalue component of $\mathbf{a}$\\
  $\xi$ \>  natural coordinate component \\
  $\eta$ \>  natural coordinate component \\
  $\omega$ \>  coefficients for modes \\
  $\bs{\omega}$ \> stacked vector of $\omega$ \\
  $\delta$  \>  model discrepancy\\
  $\sigma$ \>  variance of Gaussian random field\\
  $\phi$ \> modes\\
%  $\Lambda$ \> eigenvalue\\

{Subscripts/Superscripts}{}\\
 $rans$ \> RANS predicted results \\
 $i$ \> index of modes\\
 $obs$ \> observation \\

{Decorative symbols}{}\\
$\hat{\Box}$ \> synthetic truth \\
\end{tabbing}

%% References


%\bibliographystyle{elsarticle-num}
%\bibliography{mfu2,career,mfu}

\end{document}